\documentclass[prd,aps,showpacs,superscriptaddress,nofootinbib,preprint,tightenlines]{revtex4-1}
\pdfoutput=1
\usepackage{amsmath}
\usepackage{epsfig}
\usepackage{bm}  
\usepackage{ulem}
\newcommand{\as}{\alpha_s}
\newcommand{\td}{\text{d}}

\newcommand{\s}{\hat{s}}
\newcommand{\sa}{\hat{s}_1}

\allowdisplaybreaks

\begin{document}
\title{Inclusive Higgs Production at Large Transverse Momentum}

\author{Eric Braaten}
\email{braaten@mps.ohio-state.edu}
\affiliation{Department of Physics
                  The Ohio State University 
                   Columbus, OH 43210, USA}
                                      
\author{Hong Zhang}
\email{zhang.5676@osu.edu}
\affiliation{Department of Physics
                   The Ohio State University 
                   Columbus, OH 43210, USA}


\date{\today}
                                  
\begin{abstract}
We present a factorization formula for the inclusive production of the Higgs boson 
at large transverse momentum $P_T$ that includes all terms with the leading power of $1/P_T^2$.
The cross section is factorized into convolutions of parton distributions, 
infrared-safe hard-scattering cross sections for producing a parton, 
and fragmentation functions that give the distribution 
of the longitudinal momentum fraction of the Higgs relative to the fragmenting  parton.
The infrared-safe cross sections and the fragmentation functions
are perturbatively calculable. 
The most important fragmentation functions
are those for which the fragmenting parton is the 
top quark, gluon, $W$, $Z$, and the Higgs itself. 
We calculate the fragmentation functions at leading order in the 
Standard Model coupling constants.
The factorization formula enables the resummation of 
 large logarithms of $P_T/M_H$ due to final-state radiation 
by integrating evolution equations for the fragmentation functions.
By comparing the cross section for the process $q\bar{q}\to H t\bar t$ from the leading-power factorization formula 
at leading order in the coupling constants with the complete leading-order cross section, 
we infer that the error in the factorization formula decreases to
less than 5\% for $P_T>600$ GeV at a future $100$ TeV collider.

\end{abstract}

\maketitle

\section{Introduction}\label{sec:intro}

The discovery of the Higgs boson by the Atlas and CMS Collaborations
completed the list of elementary particles in the Standard Model of particle physics
\cite{Chatrchyan:2012ufa,Aad:2012tfa}.
It determined the only previously unknown parameter 
in the Standard Model, which is the Higgs mass  $M_H$.
It also demonstrated the relevance of the Brout-Englert-Higgs mechanism for
spontaneous breaking of the electroweak symmetry
\cite{Englert:1964et,Higgs:1964ia}.
Precise measurements of the properties of the Higgs provide strong 
constraints on new particles and forces beyond  the Standard Model.
Among the basic properties of the Higgs are its production rates in high energy collisions.
In hadron collisions, the Higgs is produced primarily with
transverse momentum $P_T$ smaller than $M_H$.
However its production rate at much larger $P_T$ is important, because it may be more sensitive to 
physics beyond the Standard Model, such as the decay of a much heavier particle into the Higgs.

The straightforward path to increasing the accuracy of theoretical predictions 
 for Higgs production in the Standard Model is complete
 higher-order perturbative calculations.
 Most phenomenologically relevant production mechanisms have been calculated with an accuracy of at least
 next-to-leading order (NLO) in the QCD coupling constant $\alpha_s$ \cite{Dittmaier:2012nh}.
 In particular, the $P_T$ distribution of the Higgs in association with a top-quark pair
 has been calculated to NLO \cite{Beenakker:2002nc}. In calculations of the Higgs $P_T$ distribution, 
 the important scales include its mass $M_H$, 
the mass of the particle from which the Higgs is emitted, 
and the center-of-mass energy {$\sqrt{\hat{s}}$ of the colliding partons, 
which may be comparable to $P_T$.
The multiscale nature of Higgs production 
 makes the complete calculation 
 at the next order in $\alpha_s$ extremely difficult.
The difficulty can be decreased by separating some of the scales.
 If $P_T$ and $\sqrt{\hat{s}}$ are smaller than the top-quark mass $M_t$,
 the scale of $M_t$ can be separated from other scales by using an effective field theory HEFT
in which virtual top-quark loops
are replaced by local interactions of the Higgs with  the vector bosons.
Several production processes have been calculated  beyond NLO using the dimension-5 operator 
in HEFT  \cite{Dittmaier:2012nh}. 
In particular, the $P_T$ distribution of the Higgs has recently been calculated to 
next-to-next-to-leading order (NNLO) \cite{Boughezal:2013uia,Chen:2014gva,Boughezal:2015dra,Boughezal:2015aha}.
Comparison with complete NLO calculations indicate that the 
difference between the exact and HEFT results  
exceeds 5\% at $P_T=150$ GeV and increases with $P_T$ \cite{Bagnaschi:2011tu}. 
The effects of the dimension-7 operators in HEFT on the Higgs $P_T$ distribution
have recently been considered \cite{Harlander:2013oja, Dawson:2014ora}.
The expansion in the higher-dimension operators of HEFT breaks down for $P_T$
above about 150~GeV.

One way to simplify calculations at  large $P_T$ and to increase their accuracy
is to separate the scale $P_T$ from the smaller scales in the problem.
In the case of inclusive hadron production at large $P_T$,
QCD factorization theorems \cite{Collins:1989gx}  imply that the leading power
in the expansion of the cross section in powers of $1/P_T^2$
comes from {\it single-parton fragmentation}:
a parton is produced with larger transverse momentum 
by a hard collision and the observed hadron is subsequently produced in the
hadronization of the hard parton into a jet.
The  leading-power (LP) factorization formula expresses the cross section as convolutions
of cross sections for producing the hard parton and
fragmentation functions $D(z,\mu)$
that give the distribution of the longitudinal  momentum fraction $z$ 
of the hadron relative to the hard parton at the momentum scale $\mu$.
All the dependence on $P_T$ is in the cross sections for producing the hard parton.
They can be calculated as expansions in the QCD coupling constant $\alpha_s$.
The fragmentation functions are nonperturbative,
but their evolution with $\mu$ is perturbative.
Their evolution equations can be used to sum large logarithms of $P_T/\Lambda_{\text{QCD}}$
in the cross sections to all orders in $\alpha_s$.

The LP factorization formula can be applied to the production
of heavy quarkonium at $P_T$ much larger than the heavy quark mass $M$.
In this case, the scale $P_T$ can be separated from the scale $M$ by including
all the dependence on $M$ in the fragmentation functions $D(z,\mu)$.
The scale $M$ can be further separated from the nonperturbative scales
in the fragmentation functions
by using NRQCD factorization \cite{Bodwin:1994jh}.
The evolution equations for the fragmentation functions
can be used to sum large logarithms of $P_T/M$ in the cross sections
to all orders in $\alpha_s$. 
The LP factorization formula with NRQCD factorization
simplifies calculations at large $P_T$
compared to fixed-order calculations with NRQCD factorization 
 by separating the scales $P_T$ and $M$,
which makes it easier to calculate higher order corrections.
The LP factorization formula can increase the accuracy compared to fixed-order calculations 
 by using evolution equations for the fragmentation functions
to sum large logarithms of $P_T/M$
to all orders in $\alpha_s$.

The factorization formulas of QCD for inclusive hadron production at large $P_T$
can be adapted straightforwardly to the production of the weak bosons 
of the Standard Model, namely $W^\pm$, $Z^0$, and Higgs.
The factorization formulas give the inclusive production rate at $P_T$
much larger than the mass $M$ of the weak boson.
At the leading power of $1/P_T$, the production mechanisms are single-parton fragmentation.
One difference from hadron production in QCD is that the weak boson itself 
must be included as one of the fragmenting partons.
The LP factorization formula simplifies calculations at large $P_T$
compared to fixed-order calculations
by separating the scales $P_T$ and $M$, which makes calculations of higher-order corrections easier.
The LP factorization formula can  increase the accuracy compared to fixed-order calculations 
by integrating evolution equations for the fragmentation functions to sum large logarithms of $P_T/M$ to all orders.
For $P_T$ of order $M_H$, there are also large threshold logarithms whose resummation 
has a significant effect on the $P_T$ distribution \cite{deFlorian:2005rr}.
In this paper, we consider only the region $P_T \gg M_H$
in which logarithms of $P_T/M_H$ may need to be resummed.

The LP factorization formula for Higgs production has been applied previously
by Dawson and Reina to the associated production of Higgs 
with a $t\bar{t}$ pair  \cite{Dawson:1997im}.
They extracted the  leading-order (LO) fragmentation function for top quark into Higgs
from the LO cross section for $q\bar q \to t \bar t H$ at large center-of-momentum energy $E_{\text{cm}}$.
They calculated an approximation to the NLO fragmentation function
using $M_H\ll M_t \ll E_{\text{cm}}$ and using a soft-gluon-emission approximation. 
They used that fragmentation function
 to estimate the NLO cross section for $t \bar t H$.
A subsequent complete NLO calculation gave results that were qualitatively compatible
with the fragmentation estimate \cite{Beenakker:2001rj}.

The LP factorization formula has recently been used to 
calculate inclusive production rates of the weak vector bosons $W^\pm$ and $Z^0$ 
at large $P_T$ \cite{Berger:2015nsa,PA-unpub}.
In Ref.~\cite{Berger:2015nsa}, this factorization formula was used to
calculate production rates at NLO in $\alpha_s$
at the Large Hadron Collider (LHC) and at a possible future 
proton-proton collider.  The leading logarithms of 
$P_T/M$, where $M$ is the mass of the vector boson,
were summed to all orders in $\alpha_s$.
The only nontrivial fragmentation function required at this order is 
that for a quark to fragment into the vector boson at leading order in 
the electroweak interaction.

In this paper, we present the LP factorization formula for inclusive
production of the Higgs at $P_T$ much larger than its mass $M_H$.
In Sec.~\ref{sec:formula}, we briefly review the QCD factorization formula 
and point out the necessary modifications to apply it to Higgs production.
In Sec.~\ref{sec:FF}, we present our results for the leading-order fragmentation functions 
for $W$, $Z$, and top quark into Higgs.
They can be identified as the fragmentation functions at appropriate initial scales.
In Sec.~\ref{sec:evolution}, we sum up the leading logarithms in the
 fragmentation functions at much higher scales by solving their evolution equations. 
 In Sec.~\ref{sec:compare}, we compare the cross section for $q \bar q \to H t \bar t$ 
from  the LP factorization formula at LO with the complete  
LO cross section in order to estimate the minimum $P_T$ above which 
the LP factorization formula is reliable.
A summary and outlook are given in Sec.~\ref{sec:conc}.


\section{Leading-Power Factorization Formula}
\label{sec:formula}

In this Section, we present the leading-power (LP) factorization formula 
for inclusive Higgs production at large $P_T$ in hadron collisions.
We compare it to the LP factorization formula for inclusive hadron production in QCD.
We describe how the LP factorization formula can simplify perturbative calculations 
by separating the scale $P_T$ from the scales of the Higgs 
mass $M_H$ and the masses of other Standard Model particles.
We also describe how the LP factorization formula can improve upon the accuracy
of perturbative calculations by summing large logarithms of $P_T$ to all orders.

\subsection{Initial-state factorization}

The differential cross section for the inclusive production of a Higgs boson
in the collision of hadrons $A$ and $B$ can be written in a factorized form \cite{Collins:1989gx}:
\begin{eqnarray}
\label{eq:Hfactorization}
\td\sigma_{AB\rightarrow H+X}(P_A,P_B,P)
&=& 
\sum_{a,b}\int_0^1\! \td x_a\, f_{a/A}(x_a,\mu) \int_0^1\! \td x_b\, f_{b/B}(x_b,\mu)
\nonumber\\
&& \hspace{1cm}
\times
\td \hat{\sigma}_{ab\to H+X}(p_a=x_aP_A,p_b=x_bP_B,P;\mu),
\end{eqnarray}
where $P$ is the momentum of the Higgs, 
$P_A$ and $P_B$ are the momenta of the colliding hadrons, 
and $p_a$ and $p_b$ are the momenta of the colliding partons.
The errors are of order $\Lambda_{\text{QCD}}^2/M_H^2$,
where $M_H$ is the mass of the Higgs.
The sums in Eq.~\eqref{eq:Hfactorization}
are over the types of QCD partons, which consist of the gluon 
and the quarks and antiquarks that are lighter than the top quark.
The integrals in Eq.~\eqref{eq:Hfactorization} are over the longitudinal momentum fractions 
of the colliding QCD partons.
The separation of the hard momentum scales of order $M_H$ and larger from the 
nonperturbative QCD momentum scale $\Lambda_{\text{QCD}}$ 
involves the introduction of an intermediate but otherwise  arbitrary
factorization scale $\mu$.
The hard-scattering cross sections $\td\hat{\sigma}$ 
 in Eq.~\eqref{eq:Hfactorization} are sensitive only to
momentum scales much larger than $\Lambda_{\text{QCD}}$, 
so they can be calculated perturbatively as expansions in powers of $\alpha_s$
and the other coupling constants of the Standard Model.
The parton distribution functions $f_{a/A}$ and $f_{b/B}$
are nonperturbative, but their evolution with $\mu$ is perturbative.
Their evolution equations can be used to sum logarithms of
$M_H^2/\Lambda_{\text{QCD}}^2$ to all orders in perturbation theory.
The colliding hadrons and the colliding partons are all treated as massless particles.
Thus the 4-momentum of the colliding parton from hadron $A$  is $p_a^\mu = x_a P_A^\mu$.
The integration range of $x_a$ is from 0 to 1,
up to kinematic constraints from  $\td\hat{\sigma}_{ab\to H+X}$.

If one or both colliding partons is a heavy (charm or bottom) quark with mass $m_Q$,
the errors in the factorization formula  in Eq.~\eqref{eq:Hfactorization}
are actually of order $m_Q^2/M_H^2$.
However, as shown by Collins, proofs of factorization can be extended to  heavy quarks \cite{Collins:1998rz}.
If the mass $m_Q$ of the heavy quark is taken into account appropriately
in the hard-scattering cross sections $\td\hat{\sigma}$, the errors in the factorization formula 
 are reduced to order $\Lambda_{\text{QCD}}^2/M_H^2$. 
The expression for the 4-momentum $p_Q$ of an incoming heavy-quark parton
 is necessarily more complicated.
The direction of the 3-momentum $\bm{p}_Q$
can be taken to coincide with the direction of the 3-momentum $\bm{P}_A$ 
of the parent hadron $A$ in the center-of-momentum frame 
of the colliding hadrons $A$ and $B$.
The magnitude of $\bm{p}_Q$ can be chosen so that
$x_Q$ is the light-front momentum fraction of the heavy quark in that frame:
\begin{equation}
\label{eq:xQ}
x_Q
= \frac{\sqrt{m_Q^2 + |\bm{p}_Q|^2} + |\bm{p}_Q|}{2  |\bm{P}_A|}
~~~~~ (\bm{P}_A + \bm{P}_B = 0).
\end{equation}
Given $x_Q$ and $|\bm{P}_A|$, this is a
linear equation for $|\bm{p}_Q|$.
It implies a lower limit on the range of integration of $x_Q$:
$x_{\textrm{min}}=m_Q/(2 |\bm{P}_A|)$.
The 4-momentum of the  heavy quark in the center-of-momentum frame 
of the colliding hadrons is
\begin{equation}
\label{eq:pQ}
p_Q^\mu  = \frac{1}{4 x_Q|\bm{P}_A|^2}
\big( (4x_Q^2 |\bm{P}_A|^2 + m_Q^2) |\bm{P}_A|, 
(4x_Q^2 |\bm{P}_A|^2 - m_Q^2) \bm{P}_A \big)^\mu
~~~~~ (\bm{P}_A + \bm{P}_B = 0).
\end{equation}
It satisfies the mass-shell constraint $p_Q^2=m_Q^2$ 
and the longitudinal momentum constraint $p_Q^0+|\bm{p}_Q|=x_Q(2|\bm{P}_A|)$.

An explicit factorization prescription named ACOT that defines hard-scattering cross sections
order-by-order in $\as$ when the colliding partons include heavy quarks
was proposed by Aivazis, Collins, Olness and Tung \cite{Aivazis:1993pi}.
Factorization theorems using the ACOT prescription were proven by Collins \cite{Collins:1998rz}.
Kramer, Olness and Soper introduced a 
simpler alternative to the ACOT prescription named S-ACOT
in which a colliding heavy-quark parton is treated as massless \cite{Kramer:2000hn}.
Three possible prescriptions for the heavy-quark mass
in the factorization formula are
\begin{itemize}
\item
 {\it zero-mass-heavy-quark} prescription.
The heavy quark mass is set to 0 in the diagrams for the
hard-scattering cross sections $\td\hat{\sigma}$. 
The 4-momentum of a colliding heavy-quark parton
from hadron $A$ is $p_Q^\mu = x P_A^\mu$ with $0<x<1$.
The errors are order $m_Q^2/Q^2$,  where $Q$ is the momentum transfer.
\item
{\it ACOT} prescription.
The heavy quark mass is $m_Q$ in the diagrams 
for the hard-scattering cross sections $\td\hat{\sigma}$. 
The 4-momentum of  a colliding heavy-quark parton is given by 
a complicated expression, such as that in Eq.~\eqref{eq:xQ}.
The errors are order $\Lambda_{\text{QCD}}^2/Q^2$.
\item
{\it S-ACOT} prescription.
The heavy-quark mass  is set to 0 
everywhere along the heavy-quark line attached to an incoming heavy-quark parton,
but the mass $m_Q$ is used for all other heavy-quark lines in a diagram
for a hard-scattering cross sections $\td\hat{\sigma}$.
The 4-momentum of a colliding heavy-quark parton from hadron $A$ is $p_Q^\mu = x P_A^\mu$ with $0<x<1$.
The errors are order $\Lambda_{\text{QCD}}^2/Q^2$.
\end{itemize} 
The S-ACOT prescription
is numerically equivalent to the ACOT prescription, even
for momentum transfers of order $m_Q$ or smaller.
 Its advantages are 
the simple expression for the 4-momentum of a colliding heavy-quark parton and
 a significant simplification
of the expressions for some of the hard-scattering amplitudes.
There are other prescriptions for taking into account the heavy quark mass
in the factorization formula in addition to the three prescriptions itemized above.
Various prescriptions for taking into account the bottom quark mass 
in the associated production 
of a Higgs and a $b \bar b$ pair were recently studied in Ref.~\cite{Bonvini:2015pxa}.

\subsection{Final-state factorization}

If the transverse momentum $P_T$ of the Higgs is much larger than its mass $M_H$, 
it is reasonable to expand the hard-scattering differential cross section 
$d\hat{\sigma}$ in Eq.~\eqref{eq:Hfactorization} in powers of $1/P_T^2$.
The leading power in $\td\hat{\sigma}/\td P_T^2$ can be determined by
dimensional analysis to be $1/P_T^4$. 
The factorization formula for the leading power can be inferred 
from the perturbative QCD factorization formula for inclusive hadron production 
at large transverse momentum \cite{Collins:1989gx}.
The leading-power (LP) factorization formula for the 
hard-scattering differential cross section 
for the inclusive production of a Higgs 
in the collision of QCD partons $a$ and $b$ is
\begin{equation}
\label{eq:LPfactorization}
\td\hat{\sigma}_{ab\rightarrow H+X}(p_a,p_b,P)
= \sum_i \int_0^1 \!\td z\, 
\td\tilde{\sigma}_{ab\rightarrow i+X}
(p_a,p_b,p_i=\tilde{P}/z;\mu) \,
D_{i\rightarrow H}(z,\mu),
\end{equation}
where $p_i$ is the momentum of the fragmenting parton 
and $\tilde{P}$ is a light-like 4-vector whose 3-vector component $\bm{P}$
 in the center-of-momentum frame of the colliding partons is the 3-momentum of the Higgs:
\begin{equation}
\label{eq:tildeP}
\tilde{P}^\mu=\big(  |\bm{P}|, \bm{P} \big)^\mu
~~~~~ (\bm{p}_a + \bm{p}_b = 0).
\end{equation}
This can be expressed in the covariant form
\begin{equation}
\label{eq:tildeP-covaraint}
\tilde{P}^\mu = P^\mu
- \frac{P\! \cdot \!(p_a+p_b) - \sqrt{[P\! \cdot \!(p_a+p_b)]^2 - M_H^2 \hat s}}{\hat s}(p_a+p_b)^\mu,
\end{equation}
where $\hat s = (p_a+p_b)^2$.
The errors in the factorization formula in Eq.~\eqref{eq:LPfactorization} are of order $M_H^2/P_T^2$.
The sum over partons $i$ is over the types of elementary particles in the Standard Model. 
The most important fragmenting partons for Higgs production  
are the weak vector bosons $W$ and $Z$, the top quark $t$,  the top antiquark $\bar t$, 
the gluon $g$, and the Higgs itself.
The integral in Eq.~\eqref{eq:LPfactorization}
is over the longitudinal momentum fraction $z$ of the Higgs 
relative to the fragmenting parton.  The integration range of $z$ is from 0 to 1,
up to kinematic constraints from $d\tilde{\sigma}_{ab\to i+X}$.
The fragmentation functions $D_{i\rightarrow H}(z,\mu)$ are distributions for $z$
 that depend on $M_H$
and on the mass  $M_i$ of the fragmenting parton.
The only dependence on $P_T$ in Eq.~\eqref{eq:LPfactorization}
is in the cross sections $\td\tilde{\sigma}$ for producing the parton $i$.
The mass $M_i$ of the fragmenting parton is set to 0 in these cross sections.
Note that the hard-scattering cross section $\td\hat{\sigma}$
on the left side of Eq.~\eqref{eq:LPfactorization}
has mass singularities in the limits $M_H\to 0$ and $M_i \to 0$.
Since the parton production cross sections $\td\tilde{\sigma}$ have no mass singularities, 
we will refer to them
as {\it infrared-safe cross sections}.
The infrared-safe cross sections $\td\tilde{\sigma}$ and the fragmentation functions 
can all be calculated perturbatively as expansions in powers of $\alpha_s$
and the other coupling constants of the Standard Model.
The separation of the hardest scale $P_T$ from the softer scales $M_H$ and $M_i$
involves the introduction of an arbitrary
factorization scale $\mu$.  We will refer to it as the {\it fragmentation scale} 
to distinguish it from the factorization scale in the 
 initial-state factorization formula in Eq.~\eqref{eq:Hfactorization}.  

The LP factorization formula in Eq.~\eqref{eq:LPfactorization}
simplifies perturbative calculations 
compared to the initial-state factorization formula in Eq.~\eqref{eq:Hfactorization}
 by separating scales in the cross section.
The fragmentation functions $D_{i\rightarrow H}$ are dimensionless functions of $z$, 
 the masses $M_H$ and $M_i$, and the fragmentation scale $\mu$.   
In the infrared-safe cross sections $\td\tilde{\sigma}_{ab\rightarrow i+X}$,
which are functions of the transverse momentum $p_T$ of the fragmenting parton,
the masses $M_H$ and $M_i$ are set equal to 0.  
The zero-mass limits are of course taken only after expressing the interactions
of the Higgs in terms of dimensionless coupling constants.
Since the only momentum scale in the infrared-safe cross sections is $p_T$,  
it is much easier to calculate higher order corrections.

The LP factorization formula in Eq.~\eqref{eq:LPfactorization}
can be used to improve upon the accuracy of fixed-order perturbative calculations
 using the initial-state factorization formula in Eq.~\eqref{eq:Hfactorization}.
Radiative corrections produce logarithms of $P_T/M_H$ and $P_T/M_i$
at higher orders in the coupling constants.
Each successive order in perturbation theory can produce an additional factor of a logarithm.
The dependence of the fragmentation functions $D_{i \to H}$ on 
the fragmentation scale $\mu$ can be expressed in terms of evolution equations 
that  have the form
\begin{eqnarray}
\label{eq:evo}
\mu^2 \frac{\partial ~}{\partial \mu^2}D_{i\to H}(z,\mu)=
\sum_{j} \int_z^1\frac{\td y}{y}P_{i\to j}(z/y,\mu)D_{j\to H}(y,\mu).
\end{eqnarray}
The sum over $j$ is over all the fragmenting partons, including the Higgs.
The functions $P_{i\to j}$
are splitting functions that can be calculated perturbatively.
If the fragmentation functions $D_{j\to H}(z,\mu_0)$ are 
 calculated at some initial scale $\mu_0$ using fixed-order perturbation theory,
the logarithms of $\mu/\mu_0$ in $D_{j\to H}(z,\mu)$
can be summed to all orders by using Eq.~\eqref{eq:evo} 
to evolve the fragmentation functions from $\mu_0$ to a larger scale $\mu$.
The logarithms of $P_T/M_H$ or $P_T/M_i$ 
in the  cross section in Eq.~\eqref{eq:LPfactorization}
can then be summed to all orders simply by
using the evolved fragmentation functions  with the fragmentation scale $\mu$  of order $P_T$. 

Since our LP factorization formula for inclusive Higgs production
in Eq.~\eqref{eq:LPfactorization}
is motivated by the corresponding factorization formula for inclusive hadron production,
we describe briefly the theoretical status of the QCD factorization formula.
The LP factorization formula for the inclusive production of a hadron
separates the scale of the transverse momentum $P_T$
of the hadron from the nonperturbative momentum scale $\Lambda_{\rm QCD}$.
In the case of a light hadron, the errors in the factorization formula 
are of order $\Lambda_{\text{QCD}}^2/P_T^2$.
The LP factorization formula for the inclusive production of a light hadron
in $e^+ e^-$ annihilation was  proven by Collins and Soper  \cite{Collins:1981uk}.
There is no apparent obstacle to extending the proof to hadron collisions  \cite{Collins:1989gx}.
In the case of a hadron that contains a single heavy quark with mass $m_Q$,
the LP factorization formula separates the scale $P_T$
from the scale $m_Q$ as well as  $\Lambda_{\text{QCD}}$.
The heavy quark mass does not present an essential complication in the proof
of factorization, but the errors in the factorization formula 
are now of order $m_Q^2/P_T^2$.
In the case of heavy quarkonium, which contains a heavy quark and antiquark,
the heavy quark mass is an essential complication.
A proof of the LP factorization formula for the production of heavy quarkonium 
was sketched by Nayak, Qiu, and Sterman \cite{Nayak:2005rt}.
The proof has been extended to the next-to-leading  power (NLP) of $1/P_T^2$
by Kang,  Ma, Qiu, and Sterman \cite{Kang:2011mg,Kang:2014tta,Kang:2014pya,Ma:2014svb}.
The NLP factorization formula has also been derived by
Fleming, Leibovich, Mehen and Rothstein using soft collinear effective theory
\cite{Fleming:2012wy,Fleming:2013qu}.
The leading corrections are suppressed by $m_Q^4/P_T^4$ 
or  $\Lambda_{\text{QCD}}^2/P_T^2$.
At NLP, there is a new production mechanism called  {\it double-parton fragmentation}:
a pair of collinear partons, such as a heavy quark and antiquark,
is  produced with larger transverse momentum by a hard collision, 
and the heavy quarkonium  is subsequently 
produced in the hadronization  of the collinear parton pair.
The LP factorization formula for inclusive Higgs production
can presumably be extended to NLP by taking into account double-parton fragmentation.

There are important differences between the factorization formula in 
Eq.~\eqref{eq:LPfactorization} for inclusive Higgs production at large $P_T$
and the analogous QCD factorization formula  for inclusive hadron production. 
One difference is that a Higgs can be produced directly in the hard scattering.
Thus the Higgs is included in the sum over fragmenting partons in Eq.~\eqref{eq:LPfactorization}.
In contrast, a hadron at large $P_T$ cannot be produced directly in the hard scattering.
Another difference is that the fragmentation functions for Higgs production are completely perturbative.
They can be calculated order-by-order in the Standard Model coupling constants. 
In contrast, the fragmentation functions for hadron production are nonperturbative,
although their evolution with the  fragmentation scale is perturbative.

\subsection{Top-quark mass}
\label{sec:topmass}

The top quark is the only particle in the Standard Model whose mass 
is larger than that of the Higgs.
Those terms in the LP factorization formula in Eq.~\eqref{eq:LPfactorization}
for which the infrared-safe cross section $\td\tilde{\sigma}$
has a top-quark in the final state
have fractional errors that are of order $M_t^2/P_T^2$ instead of order $M_H^2/P_T^2$.
The errors can be decreased to order $M_H^2/P_T^2$
by taking into account $M_t$ in the infrared-safe cross section $\td\tilde{\sigma}$.
Infrared-safe cross sections $\td\tilde{\sigma}$ that depend on $M_t$
but have no mass singularities as $M_t \to 0$
can be constructed order-by-order in the coupling constants
by using the same strategy as in the ACOT scheme for
taking into account the heavy-quark mass in initial-state factorization \cite{Aivazis:1993pi}.

If the top-quark mass is not neglected,
the relation between the 4-momenta $P$ of the Higgs and 
$p$ of the fragmenting top quark is necessarily more complicated than 
the expression $p=\tilde{P}/z$ in Eq.~\eqref{eq:LPfactorization}.
The direction of the 3-momentum $\bm{p}$
can be taken to coincide with that of the Higgs 3-momentum $\bm{P}$ in 
the center-of-momentum frame of the colliding partons $a$ and $b$.
The magnitude of $\bm{p}$ can be chosen so that
$z$ is the light-front momentum fraction of the Higgs in that frame:
\begin{equation}
\label{eq:zlc}
z  = \frac{\sqrt{M_H^2+|\bm{P}|^2} + |\bm{P}|}{\sqrt{M_t^2+|\bm{p}|^2} + |\bm{p}|}
~~~~~ (\bm{p}_a + \bm{p}_b = 0).
\end{equation}
Given $z$ and $|\bm{P}|$, this is a
linear equation for $|\bm{p}|$.
It implies an upper limit on $z$
that is less than 1 if $|\bm{P}| <(M_t^2-M_H^2)/(2M_t):$
\begin{equation}
\label{eq:zmax}
z_{\textrm{max}} = \min\left(1,\Big(\sqrt{M_H^2+|\bm{P}|^2} + |\bm{P}|\Big)/M_t\right).
\end{equation}
The 4-momentum of the  top quark in the center-of-momentum frame 
of the colliding partons is
\begin{eqnarray}
\label{eq:pt}
p^\mu  = \frac{1}{2z}
\left( \frac{ \big( \sqrt{|\bm{P}|^2+ M_H^2} + |\bm{P}| \big)^2 + z^2 M_t^2 }
               {\sqrt{|\bm{P}|^2+ M_H^2} + |\bm{P}| } , 
\frac{ \big( \sqrt{|\bm{P}|^2+ M_H^2} + |\bm{P}| \big)^2 - z^2 M_t^2 }
               {\big(\sqrt{|\bm{P}|^2+ M_H^2 } + |\bm{P}|\big) |\bm{P}|}  \bm{P} \right)^\mu
\nonumber\\
(\bm{p}_a + \bm{p}_b = 0). \hspace{0.5cm}
\end{eqnarray}
This satisfies the mass-shell constraint $p^2 = M_t^2$
and the longitudinal momentum constraint
$P^0 +  |\bm{P}| = z (p^0 +|\bm{p}|)$.

One can obtain the  improved accuracy of the factorization formula with a massive top quark
and some of the simplifications of the factorization formula with a zero-mass top quark
by using a  hybrid factorization prescription
analogous to the S-ACOT prescription for the heavy-quark mass 
in the initial state factorization formula  \cite{Kramer:2000hn}. In this hybrid prescription,
the mass of the top quark is set to zero in an infrared-safe cross section 
if that top quark
is the fragmenting parton.
Three possible prescriptions for the top-quark mass
in the factorization formula are
\begin{itemize}
\item
{\it zero-mass-top-quark (ZMTQ)} prescription.
The top-quark mass is set to 0 in the diagrams for the
infrared-safe cross sections $\td\tilde{\sigma}$. 
The 4-momentum of a fragmenting top quark is $p^\mu = \tilde{P}^\mu/z$
with $0<z<1$. The errors are order $M_t^2/P_T^2$.
\item
{\it massive-top-quark (MTQ)} prescription.
The top-quark mass is $M_t$ in the diagrams 
for the infrared-safe cross sections $\td\tilde{\sigma}$.
This prescription is the analog of the ACOT prescription for initial-state factorization.
The 4-momentum of the fragmenting top quark is given by 
a complicated expression, such as that in Eq.~\eqref{eq:pt}.
The errors are order $M_H^2/P_T^2$. 
\item
{\it hybrid} prescription.
The  top-quark mass  is set to 0 
everywhere along the top-quark line attached to a fragmenting top quark,
but the mass $M_t$ is used for all other top-quark lines in a diagram
for an infrared-safe cross sections $\td\tilde{\sigma}$.
This prescription is the analog of the S-ACOT prescription for initial-state factorization.
The 4-momentum of a fragmenting top quark is  $p^\mu = \tilde{P}^\mu/z$ with $0<z<1$. 
The errors are  order $M_H^2/P_T^2$. 
\end{itemize}
Based on experience with the heavy-quark mass in initial-state factorization,
the hybrid prescription can be expected to  
be numerically equivalent to the MTQ prescription for $P_T \gg M_H$.
Its advantages are the simple expression for the 4-momentum of the fragmenting top quark
and significant simplifications in some of the infrared-safe cross sections.

The issue of how the mass of a fragmenting parton should be taken into account 
arises also in applications of the LP factorization formula to the inclusive production  
of charm or bottom hadrons at large $P_T$.
The effects of prescriptions for taking into account the charm quark mass 
on the inclusive production of charm hadrons at large $P_T$ 
at NLO in $\alpha_s$ has been studied in Ref.~\cite{Kneesch:2007ey}.
The masses analogous to $M_t$ and $M_H$ 
are the mass of the charm quark and the mass of the charm hadron, respectively.
For $e^+ e^-$ annihilation at the $Z^0$ resonance,
the effects of the masses are negligible.
For $e^+ e^-$ annihilation in the $b \bar b$ threshold region,
the effects of the charm hadron mass are appreciable,
but  the effects of the charm quark mass are less important.


\section{Leading-order Fragmentation functions}
\label{sec:FF}

In this section, we present the most important fragmentation functions for Higgs production in the Standard Model at leading order in the strong, electroweak, and Yukawa coupling constants.
The most important fragmenting partons are the Higgs itself, the 
weak vector bosons $W$ and $Z$, the top quark, and the gluon.

\subsection{Diagrammatic calculation}

\noindent

Fragmentation functions can be defined to all orders in the coupling constants
in terms of matrix elements of composite operators 
in a quantum field theory \cite{Collins:1981uw}.
The composite operators consist of a local  source operator and a local sink operator
that are connected by an eikonal factor that may also be an operator.
The source and sink operators and the eikonal factor 
depend on a light-like 4-vector $n$.
The fragmentation functions can be calculated diagramatically 
using Feynman rules introduced by Collins and Soper \cite{Collins:1981uw}.
The Feynman rules are summarized in Appendix~\ref{app:FRules}.
The diagrams have a source vertex that creates one or more virtual partons 
and a sink  vertex that annihilates virtual partons.
The source and sink vertices are connected by an eikonal line.
The momentum $K$ flowing from the source into the virtual partons and the eikonal line can be interpreted
as the momentum of the fragmenting  parton.
There is a cut through the diagram that separates the source and the sink,
passes through the eikonal line, and also cuts other lines. 
The other cut lines can be interpreted as a final state 
from the fragmentation of the virtual parton.  
The cut lines include a Higgs that is on its mass shell with a specified momentum $P$.  
The longitudinal momentum fraction of the Higgs is defined
by $z = P\cdot n/K \cdot n$. 
The expression for a diagram involves integrals over the phase space of 
 the cut lines other than the Higgs. 
The expression may also involve integrals over loop momenta.
Ultraviolet divergences in the phase space integrals and the loop
 integrals are cancelled by renormalization
of the composite operator and by the conventional renormalization of the quantum field theory.

The leading-order (LO) contribution to most fragmentation functions for Higgs production
comes from a tree-level process $i^* \to H+i$,
where the asterisk indicates that the fragmenting parton is a virtual particle.
The diagram for the LO fragmentation function $D_{i \to H}(z,\mu)$
is a cut diagram with a tree-level diagram on each side of the cut.  
The source operator creates a single virtual particle $i^*$.
The cut lines are the Higgs and the on-shell particle $i$.
The integral over the phase space of particle $i$
can be reduced to an integral over the invariant mass squared $t$
of  $H+i$.
The integral over $t$ is ultraviolet divergent.  The definition of the factorization prescription
must include a prescription for removing that divergence.
We will consider two prescriptions for removing the ultraviolet divergence:
\begin{itemize}
\item
The {\it $\overline{\text{MS}}$ factorization scheme} is defined 
to all orders in perturbation theory by dimensional regularization 
and modified minimal subtraction. This prescription introduces an arbitrary fragmentation scale $\mu$ through multiplicative factors of  $\mu^{4-D}$,
where $D$ is the number of space-time dimensions. The subtraction
followed by the limit $D \to 4$ results in logarithmic dependence 
on $\mu$. 
\item
The {\it invariant-mass-cutoff (IMC) factorization scheme} is defined at LO in perturbation theory
by imposing an upper limit $t < \mu^2$ on the integral over the invariant mass
of the Higgs and the additional particle.
\end{itemize}
The IMC factorization scheme may only be applicable at  LO.
The obstacles  to extending  it to higher orders are that 
the invariant-mass cutoff is not sufficient to remove ultraviolet divergences beyond LO 
and that the invariant mass is not gauge invariant beyond LO.

In the IMC scheme,
the fragmentation scale $\mu$ has a simple physical interpretation
as the maximum invariant mass of the jet that includes the Higgs.
For the tree-level fragmentation process $i^* \to H + i$,
there is a lower limit on the invariant mass of the final-state particles:
$t > M_H^2/z + M_i^2/(1-z)$.  The minimum 
invariant mass is $\mu_{0,i} = M_H+M_i$, which occurs at $z = M_H/(M_H+M_i)$.  
In the $\overline{\text{MS}}$ scheme, the fragmentation scale $\mu$ 
does not have any direct physical interpretation.
By comparing LO fragmentation functions in the $\overline{\text{MS}}$ scheme
and the  IMC scheme in the limit $\mu \gg M_H$, 
the fragmentation scale $\mu$ in the $\overline{\text{MS}}$ scheme
can be related to the maximum invariant mass of the jet that includes the Higgs.

Radiative corrections to the fragmentation functions produce logarithms of 
$\mu/M_H$ at higher orders in the coupling constants.
Each successive order in perturbation theory can produce an additional factor of a logarithm.
The leading logarithms of $\mu/M_H$ from higher orders in perturbation theory 
can be summed to all orders by solving the evolution equation in Eq.~\eqref{eq:evo}
with the boundary condition that the fragmentation function at some initial scale $\mu_0$ of order $M_H$ 
is equal to the fixed-order fragmentation function at that scale.
The evolution of the fragmentation functions will be discussed in 
Section~\ref{sec:evolution}.

We will find below that there are large differences between fragmentation functions in the 
$\overline{\text{MS}}$  scheme and the IMC scheme.
However, when we add the terms in the LP factorization formula in Eq.~\eqref{eq:LPfactorization},   
the differences in a hard-scattering cross section are suppressed by
 a factor  of $M_H^2/P_T^2$.
This will be illustrated in Sec.~\ref{sec:compare}, 
 where we compare the contributions from the subprocess $q\bar{q}\to H t \bar t$ 
 to the LP factorization formula at LO and to the complete LO  cross section 
for inclusive Higgs production at a 100~TeV $pp$ collider.

The invariant-mass-cutoff scheme 
for the fragmentation functions of light quarks into a weak vector bosons
was used by Berger et al.\ in
Ref.~\cite{Berger:2015nsa}  to sum the leading logarithms of  $P_T/M_V$
in the cross sections for the production of $W^\pm$ and $Z_0$ at NLO in $\alpha_s$.
Comparing with fixed NLO predictions, the resummed NLO predictions 
show a moderate reduction of the theoretical uncertainly 
and an increase in the $P_T$ distribution 
by about $5\%$ at  $P_T\gtrsim 500$~GeV.

\subsection{Higgs fragmentation}

Perhaps the most important parton that fragments into the Higgs is the Higgs boson itself.
The LO fragmentation function for Higgs 
into Higgs is a delta function:
\begin{equation}
\label{eq:DHH}
D_{H\to H}(z)=\delta(1-z) + {\cal O}(g_W^2, y_t^2).
\end{equation}
The leading corrections  
are of order $g_W^2$ from the coupling of the Higgs to weak vector bosons 
and of order $y_t^2$ from the Yukawa coupling of the Higgs to the top quark. 

If we insert the LO  fragmentation function for $H\to H$
in Eq.~(\ref{eq:DHH}) into the factorization formula in Eq.~\eqref{eq:LPfactorization},
the $i=H$ term reduces to 
$\td\tilde{\sigma}_{ab\rightarrow H+X} (p_a,p_b,p_H=\tilde{P};\mu)$,
 which corresponds to the direct production of 
a massless Higgs in the hard scattering.
The infrared-safe cross section  
$\td\tilde{\sigma}_{ab\rightarrow H+X}$ can be obtained
from the hard-scattering cross section 
$\td\hat{\sigma}_{ab\rightarrow H+X}$ defined by the initial-state factorization formula 
in Eq.~\eqref{eq:Hfactorization} by  first subtracting mass singularities 
and then setting the Higgs mass and the masses of other particles to 0.
This procedure of course offers no simplifications over  the complete LO
calculation of the  hard-scattering cross section $\td\hat{\sigma}_{ab\rightarrow H+X}$.
However it is possible to obtain the infrared-safe cross section  
$\td\tilde{\sigma}_{ab\rightarrow H+X}$ through a much simpler calculation
that involves only the scale $p_T$.
One would start with the hard-scattering cross section 
$\td\hat{\sigma}_{ab\rightarrow H+X}$ with all masses set to 0
and with the mass singularities  dimensionally regularized.
One would subtract
the mass singularities, and then
finally take the limit $D \to 4$ to get $\td\tilde{\sigma}_{ab\rightarrow H+X}$.

If we insert the LO fragmentation function for $H\to H$
in Eq.~(\ref{eq:DHH}) into the evolution equation in Eq.~(\ref{eq:evo}),
the $j=H$ term reduces to the inhomogeneous term $P_{i\to H}(z,\mu)$.
All the other terms are 
 homogeneous in the fragmentation functions.
If the only relevant term in the renormalization of the composite operator
is the convolution of the $i\to H$ splitting kernel with the LO fragmentation function for $H\to H$, 
we can determine the LO splitting function $P_{i \to H}$ for $i \ne H$
simply by differentiating the LO fragmentation function:
\begin{equation}
\label{eq:PiH}
P^{\rm LO}_{i \to H}(z, \mu) = 
\mu^2 \frac{\partial~}{\partial \mu^2} D^{\rm LO}_{i\to H}(z,\mu).
\end{equation}
This relation applies to all the fragmentation functions calculated in this paper.

\subsection{Weak vector boson fragmentation}
\label{subsec:WeakFF}
 
The weak vector bosons in the Standard Model are $W^+$, $W^-$, and $Z^0$.
In the cross section for producing a weak vector boson $V$, 
the leading power of $1/P_T$ comes from a  transversely-polarized $V$. 
The Feynman rules for a $V$ fragmentation function are 
described in Appendix~\ref{app:FRules}.
The leading-order contribution to the fragmentation function for $V$
into Higgs  comes from the tree-level process $V^* \to H+V$.
The cut diagram is shown in Fig.~\ref{fig:V->H_FD}.
It can be expressed as an integral over the invariant mass squared $t$
of the final state $H+V$ from $M_H^2/z + M_V^2/(1-z)$ to $\infty$.
The minimum invariant mass is $\mu_{0,V} = M_H + M_V$.

\begin{figure}
\begin{center}
\includegraphics[width=0.3\textwidth]{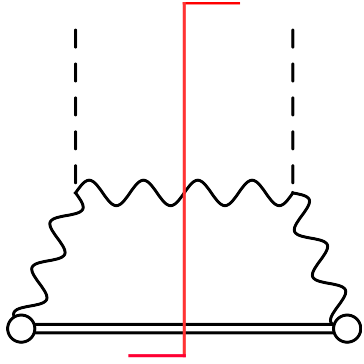}
\caption{Feynman diagram for the $V\to H$ fragmentation function at LO.
The open circles are vertices for the source and sink operators.
The double line connecting them represents the eikonal factor.
\label{fig:V->H_FD}}
\end{center}
\end{figure}

The LO fragmentation function for  $V$ into Higgs 
in the $\overline{\text{MS}}$ factorization scheme is
\begin{eqnarray}
\label{eq:V->H}
D_{V \to H}(z,\mu) = 
\frac{y_V^2}{8\, \pi^2} z\,(1-z)
\Big\{ \log\frac{\mu^2}{M_V^2}  - \log[z^2+4\,\zeta_V(1-z)]
+\frac{2}{z^2+4\,\zeta_V\,(1-z)}\,\Big\},
\end{eqnarray}
where $\zeta_V\equiv M_H^2/(4M_V^2)$ and $y_V = M_V/v$.
The numerical values of the mass-squared ratios are $\zeta_W = 0.611$
and $\zeta_Z = 0.475$. 
The numerical values of the coupling constants are $y_W=0.327$
and $y_Z = 0.371$.
By differentiating Eq.~\eqref{eq:V->H} with respect to $\mu^2$ as in Eq.~\eqref{eq:PiH},
we obtain the 
LO splitting function for $V \to H$:
\begin{equation}\label{eq:PV->H}
P_{V\to H}(z) = \frac{y_V^2} {8\, \pi^2}z\,(1-z).
\end{equation}

The LO fragmentation function for $V$ into Higgs 
in the IMC  factorization scheme is
\begin{equation}\label{eq:V->Hcut}
D_{V \to H}^{\text{IMC}}(z,\mu) = 
\int_{0}^{\mu^2} \frac{\td t}{t}d_{V \to H}(z,t),
\end{equation}
where the integrand is
\begin{equation}
\label{eq:dV->H}
d_{V \to H}(z,t) = \frac{y_V^2} {8\, \pi^2} \,
\left( \frac{z\,(1-z)\,t}{t-M_V^2} + \frac{[2-z^2-4\,\zeta_V\,(1-z)] M_V^2t }{(t-M_V^2)^2}\right)~
\theta\!\left(t-\frac{M_H^2}{z}-\frac{M_V^2}{1-z}\right).
\end{equation}
The $\theta$ function
provides the lower limit on the integral over $t$ in Eq.~\eqref{eq:V->Hcut}.
Evaluating the integral,   
we obtain an analytic expression for the fragmentation 
 function:
\begin{eqnarray}
\label{eq:V->Hcutoff}
D_{V\to H}^{\text{IMC}}(z,\mu) &=&
 \frac{y_V^2} {8\, \pi^2}\,z\,(1-z)\bigg\{
 \log\frac{\mu^2-M_V^2}{M_V^2}
+\log[z(1-z)]
 - \log[z^2+4\,\zeta_V(1-z)]
\nonumber\\
&& \hspace{0cm}
+ [2-z^2-4\,\zeta_V\,(1-z)]
\left[ \frac{1}{z^2+4\,\zeta_V (1-z)} - \frac{M_V^2}{z\,(1-z)(\mu^2-M_V^2)} \right]
\bigg\}
\nonumber\\
&&  \hspace{0cm}
\times
\theta\!\left(\mu^2-\frac{M_H^2}{z}-\frac{M_V^2}{1-z}\right).
\end{eqnarray}
By differentiating Eq.~\eqref{eq:V->Hcut}
with respect to $\mu^2$ as in Eq.~\eqref{eq:PiH},
we obtain the LO splitting function for $V \to H$:
\begin{equation}\label{eq:PV->Hcut}
P_{V\to H}^{\text{IMC}}(z,\mu)=d_{V\to H}(z,\mu^2).
\end{equation}
For $\mu\gg M_V$,
we can drop terms in $D_{V\to H}^{\text{IMC}}(z,\mu)$ that are suppressed by $M_V^2/\mu^2$.
In this limit, the difference between the fragmentation functions 
in the IMC scheme in Eq.~\eqref{eq:V->Hcutoff}
and in the $\overline{\text{MS}}$ scheme in Eq.~\eqref{eq:V->H} is simple: 
\begin{equation}\label{eq:DV->Hdiff}
D_{V\to H}^{\text{IMC}}(z,\mu)  \approx D_{V\to H}(z,\mu)
+ \frac{y_V^2} {8\, \pi^2} z(1-z) \left( \log[z(1-z)] - 1 \right).
\end{equation}
The additional term can be absorbed into $D_{V\to H}(z,\mu)$
by making the substitution $\mu^2 \to e^{-1}\, z(1-z) \mu^2$.
Thus the fragmentation function in the IMC scheme
is approximately equal to the fragmentation function 
in the $\overline{\text{MS}}$ scheme with  a $z$-dependent factorization scale.

\begin{figure}[htz]
\begin{center}
\includegraphics[width=0.72\textwidth]{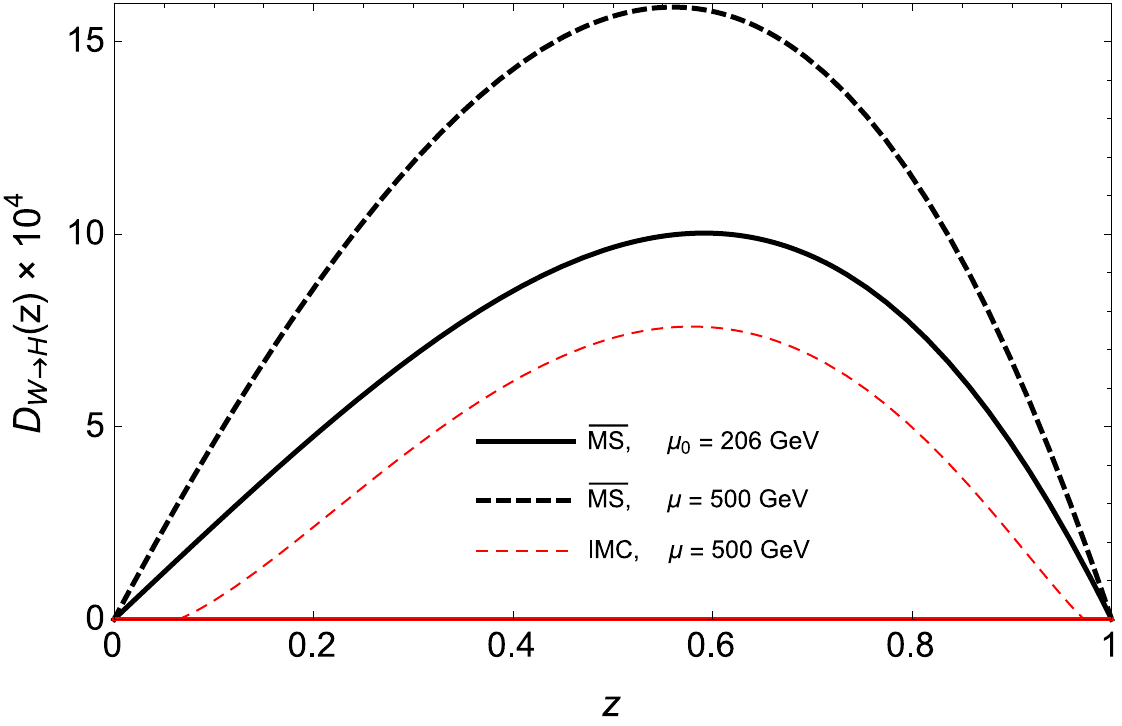}
\includegraphics[width=0.72\textwidth]{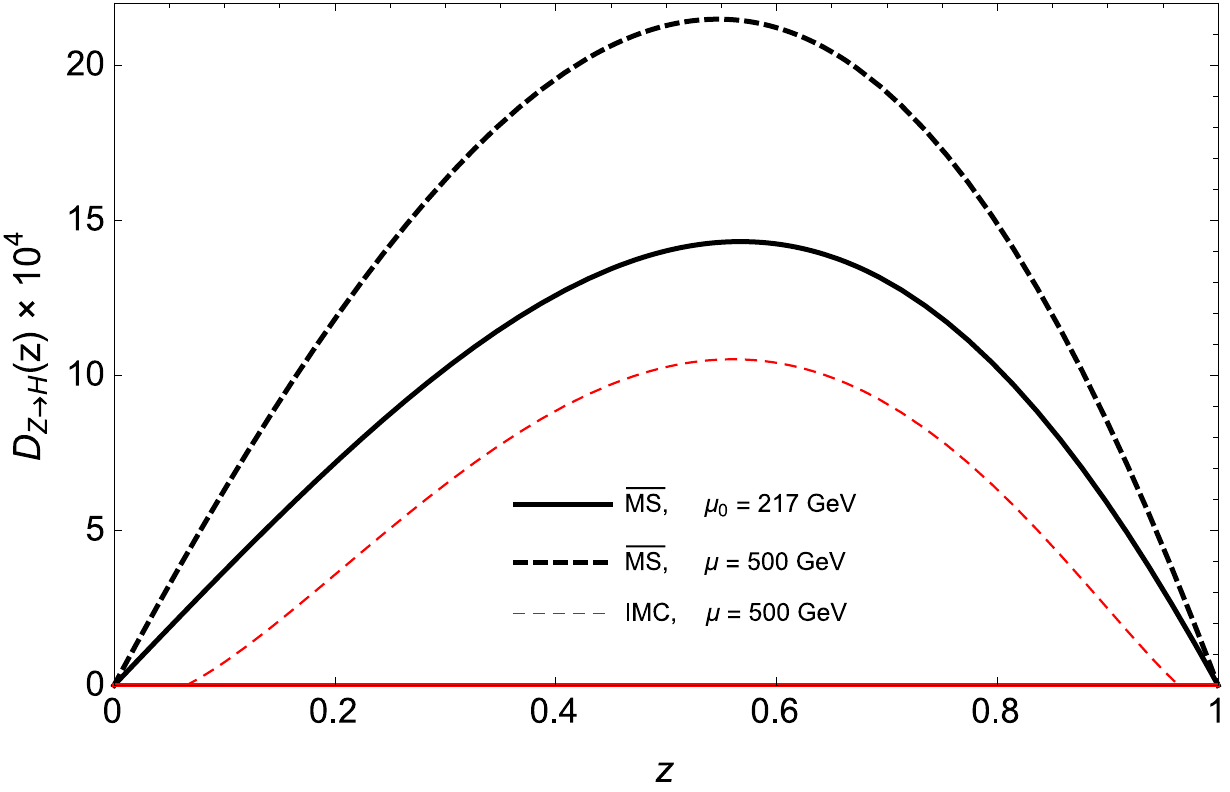}
\caption{Fragmentation functions
for $W^\pm$ into Higgs and $Z^0$ into Higgs at LO. 
The fragmentation functions in the $\overline{\text{MS}}$ scheme are shown at 
the scales $\mu_{0,V}=M_H+M_V$ (solid curves) and $500$~GeV (thicker dashed curves). 
The fragmentation functions in the IMC scheme, which are equal to zero for all $z$  
at the scale $\mu_{0,V}$, are shown at the scale $500$~GeV (thinner dashed curves).
\label{fig:V->H}}
\end{center}
\end{figure}

The LO fragmentation functions for $W^\pm$ into Higgs and $Z^0$ into Higgs
are illustrated in Figure~\ref{fig:V->H}.
The LO fragmentation functions in the $\overline{\text{MS}}$ scheme 
and in the IMC scheme
are shown at the initial scale $\mu_{0,V}=M_H+M_V$, 
which is 206~MeV for $W^\pm$ and 217~GeV for $Z^0$,
and at $\mu = 500$~GeV.
In the $\overline{\text{MS}}$ scheme, 
the fragmentation functions are positive and they vanish at the endpoints $z=0,1$.
The initial fragmentation function at $\mu = \mu_{0,V}$ has its maximum  at $z=0.59$ for $W \to H$
and at $z=0.57$ for $Z \to H$.
As $\mu$ increases, the fragmentation function increases 
and the position of its maximum shifts downward, asymptotically approaching $z = 0.5$.
In the IMC scheme, the initial fragmentation function at $\mu_{0,V}$  is 0.
At larger scales $\mu$, it is nonzero only in the subinterval $(z_-,z_+)$, where
\begin{equation}\label{eq:z+-}
z_{\pm} = 
\frac{\mu^2-M_t^2+M_H^2 \pm \big[ (\mu^2-M_t^2-M_H^2) - 4 M_t^2 M_H^2 \big]^{1/2}}{\mu^2}.
\end{equation}
For $\mu$ just above $\mu_{0,V}$, the fragmentation function is nonzero only 
for $z$ near 0.61 for $W \to H$ and near 0.58 for $Z\to H$.
As $\mu$ increases, the interval $(z_-,z_+)$ expands towards its asymptotic limit (0,1).
The fragmentation functions in the IMC scheme remain well below those in the $\overline{\text{MS}}$ scheme
until $\mu$ is very large.
In the central region of $z$, the IMC fragmentation functions  at the scale $\mu = 500$~GeV
are  still below
 those for the $\overline{\text{MS}}$ fragmentation functions at the initial scale $\mu_{0,V}$.
They are smaller than the  $\overline{\text{MS}}$ fragmentation functions at $\mu = 500$~GeV
by about a factor of 2.
\subsection{Top-quark fragmentation}

The leading-order contribution to the fragmentation function for top quark  
into Higgs  comes from the tree-level process $t^* \to H+t$. 
The Feynman diagram is shown in Fig.~\ref{fig:t->H_FD}.
The Feynman rules for a $t$ fragmentation function are 
described in Appendix~\ref{app:FRules}.
The fragmentation function
can be expressed as an integral over the square of the  invariant mass 
of the final state $H+t$ from $M_H^2/z + M_t^2/(1-z)$ to $\infty$.
The minimum invariant mass is $\mu_{0,t} = M_t+M_H$.

\begin{figure}
\begin{center}
\includegraphics[width=0.3\textwidth]{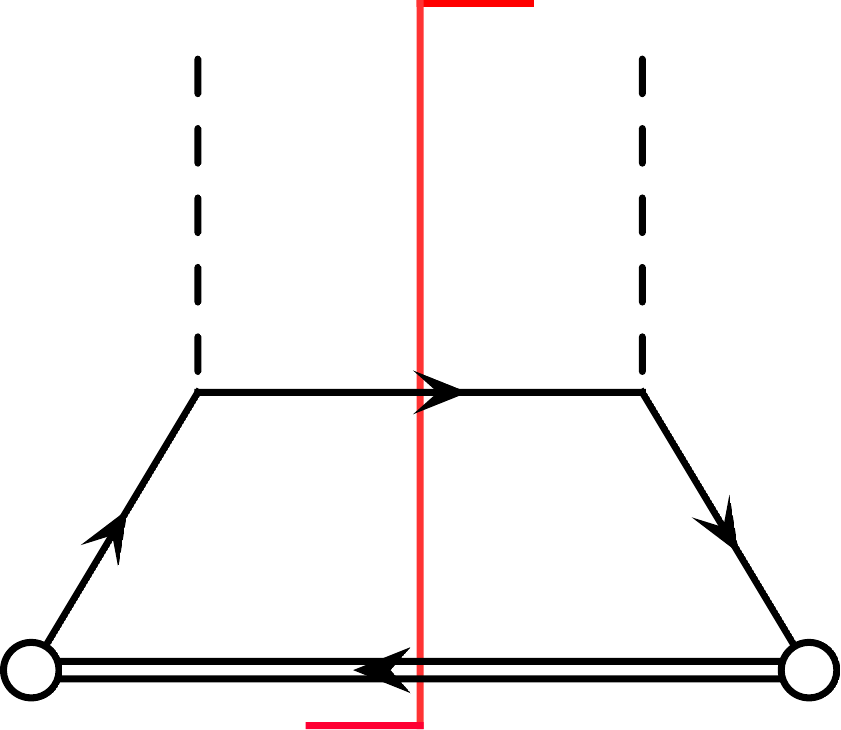}
\caption{Feynman diagram for the $t\to H$ fragmentation function at LO.
\label{fig:t->H_FD}}
\end{center}
\end{figure}

The LO fragmentation function for $t \to H$
in the $\overline{\text{MS}}$ factorization scheme is
\begin{equation}\label{eq:t->H}
D_{t\to H}(z,\mu) =
\frac{y_t^2}{16 \pi^2} \, z
\left( \log\frac{\mu^2}{M_t^2}
- \log\big[z^2+4\,\zeta_t(1-z)\big]
+4\,(1-\zeta_t)\frac{1-z}{z^2+4\,\zeta_{t}\, (1-z)}
\right),
\end{equation}
where $\zeta_t \equiv M_H^2/(4M_t^2) \approx 0.13$.
The fragmentation function for $\bar{t} \to H$  is the same.
By differentiating with respect to $\mu^2$
as in Eq.~\eqref{eq:PiH},
we obtain the LO splitting function for $t \to H$:
\begin{equation}
\label{P:t->H}
P_{t\to H}(z) = \frac{y_t^2}{16 \pi^2} \,z.
\end{equation}

The LO fragmentation function for $t\to H$  
in the IMC factorization scheme is
\begin{equation}
\label{eq:t->Hcutint}
D_{t \to H}^{\text{IMC}}(z,\mu) = 
\int_{0}^{\mu^2} \frac{\td t}{t}d_{t \to H}(z,t),
\end{equation}
where the integrand is
\begin{equation}
\label{eq:dt->H}
d_{t \to H}(z,t) = \frac{y_t^2}{16 \pi^2}
\left\{ \frac{z\, t}{t-M_t^2} +4(1-\zeta_t)\frac{M_t^2 \, t}{(t-M_t^2)^2}\right\}\theta\left(t-\frac{M_H^2}{z}-\frac{M_t^2}{1-z}\right).
\end{equation}
The $\theta$ function
provides the lower limit on the integral over $t$ in Eq.~\eqref{eq:t->Hcutint}.
Evaluating the integral,  
we obtain an analytic expression for the fragmentation function:
\begin{eqnarray}
\label{eq:t->Hcut}
D_{t\to H}^{\text{IMC}}(z,\mu) &=&
\frac{y_t^2}{16 \pi^2} z\,
\bigg\{ \log\frac{\mu^2-M_t^2}{M_t^2} 
+\log[{z(1-z)}] -\log[z^2+4\,\zeta_t (1-z)]
\nonumber \\
&& \hspace{0cm}
+4\,(1-\zeta_t)
\left[\frac{1-z}{z^2+4\,\zeta_t(1-z)}-\frac{M_t^2}{z\,(\mu^2-M_t^2)}\right]
\bigg\}\theta\left(\mu^2-\frac{M_H^2}{z}-\frac{M_t^2}{1-z}\right).
\end{eqnarray}
By differentiating Eq.~\eqref{eq:t->Hcutint} with respect to $\mu^2$
as in Eq.~\eqref{eq:PiH},
we obtain the LO splitting function for $t \to H$:
\begin{equation}\label{eq:Pt->Hcut}
P_{t\to H}^{\text{IMC}}(z,\mu)= d_{t \to H}(z,\mu^2).
\end{equation}
For $\mu\gg M_t$, 
 we can drop terms in $D_{t\to H}^{\text{IMC}}(z,\mu)$ that are suppressed by $M_t^2/\mu^2$. 
In this limit, the difference between the fragmentation functions 
in the IMC scheme in Eq.~\eqref{eq:t->Hcut}
and in the $\overline{\text{MS}}$ scheme in Eq.~\eqref{eq:t->H} is simple:
\begin{equation}\label{eq:Dt->Hdiff}
D_{t\to H}^{\text{IMC}}(z,\mu)  \approx D_{t\to H}(z,\mu)
+ \frac{y_t^2}{16 \pi^2} \,z \log[z(1-z)].
\end{equation}
The additional term can be absorbed into $D_{t\to H}(z,\mu)$
by making the substitution $\mu^2 \to  z(1-z) \mu^2$.
Thus the fragmentation function in the IMC scheme
is approximately equal to the fragmentation function 
in the $\overline{\text{MS}}$ scheme with  a $z$-dependent fragmentation scale.

\begin{figure}
\begin{center}
\includegraphics[width=0.82\textwidth]{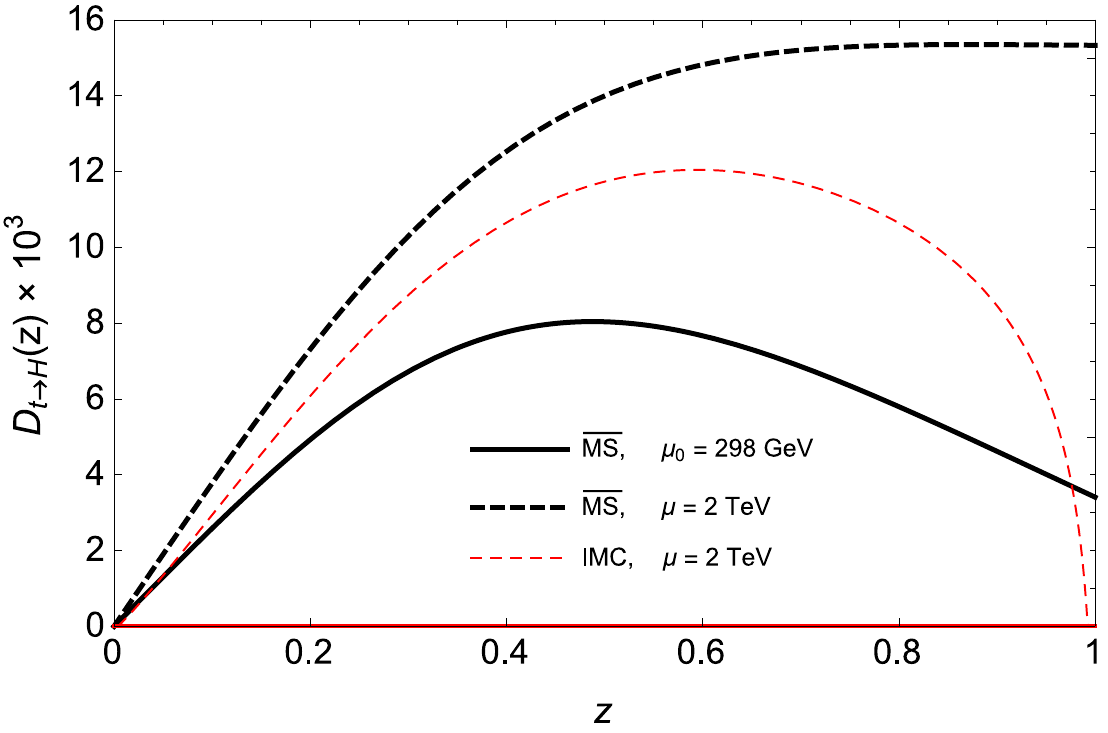}
\caption{Fragmentation functions for top quark into Higgs at LO. 
The fragmentation function in the $\overline{\text{MS}}$ scheme is shown 
at the scale $\mu_{0,t}=M_t+M_H$ (solid curve) and $2$~TeV (thicker dashed curve). 
The fragmentation function in the IMC scheme,
which is equal to zero for all $z$  at the scale  $\mu_{0,t}$, 
is shown at the scale $2$~TeV (thinner dashed curve).
\label{fig:t->H}}
\end{center}
\end{figure}

The LO fragmentation functions for $t$ into Higgs
are illustrated in Fig.~\ref{fig:t->H}.
The fragmentation functions in the $\overline{\text{MS}}$ scheme and in the IMC scheme
are shown at the initial scale
$\mu_{0,t}=M_t+M_H$, which is 298~MeV, and at $\mu=2$~TeV.
In the $\overline{\text{MS}}$ scheme, 
the fragmentation function is positive, and it vanishes at $z=0$.
At the initial scale $\mu_{0,t}$, it has a maximum at $z=0.49$.
As $\mu$ increases, the  fragmentation function increases and the position of its maximum
moves to larger $z$, reaching $z=1$ at $\mu=2.06$~TeV.
It remains at $z=1$ for larger $z$.
In the IMC scheme, the initial fragmentation function
at $\mu_{0,t}$ is 0.  
At larger scales $\mu$, it is nonzero only in a subinterval $(z_-,z_+)$ of (0,1).
For $\mu$ just above $\mu_{0,V}$, the fragmentation function is nonzero only 
for $z$ near 0.42.
As $\mu$ increases, the interval $(z_-,z_+)$ expands towards its asymptotic limit (0,1).
At $\mu=2$~TeV, the IMC fragmentation function is a little lower than
the $\overline{\text{MS}}$ fragmentation function in the central region of $z$, 
 but it is much lower near $z=1$, where it vanishes.

In Ref.~\cite{Dawson:1997im}, Dawson and Reina deduced the fragmentation function 
for $t \to H$ from the energy distribution of the Higgs in the process
$q \bar q \to t \bar t H$.
The longitudinal energy fraction at leading order is $z = 2 E/\sqrt{s}$,
where $E$ is the energy of the Higgs in the center-of-momentum frame 
and $\sqrt{s}$ is the center-of-mass energy of the colliding $q \bar q$.
They expressed the LO differential energy distribution as
\begin{equation}
\label{sig-ttH}
d \sigma [ q \bar{q} \to t \bar{t} H] =  2\, \sigma [ q\bar{q} \to t \bar{t}]~
D_{t\to H}(z)\, dz.
\end{equation}
Taking the limit $M_H^2 \ll M_t^2 \ll s$, they obtained the LO fragmentation function
\begin{equation}
\label{D-DR}
D_{t\to H}(z)=
\frac{y_t^2}{16 \pi^2}
\left( z \log\frac{(1-z)s}{M_t^2}
+\frac{4(1-z)}{z} \right).
\end{equation}
This can be obtained from the LO fragmentation function 
in the $\overline{\text{MS}}$ factorization scheme in Eq.~\eqref{eq:t->H}
by setting $\zeta_t = 0$ and by making the substitution
$\mu^2 \to z^2(1-z) s$.
Thus the fragmentation function in Eq.~\eqref{D-DR} can be 
 approximated by the $\overline{\text{MS}}$ fragmentation function
with a $z$-dependent renormalization scale. 
Dawson and Reina also calculated an approximation to the NLO fragmentation 
function at small values of $z$ \cite{Dawson:1997im}.
It would be interesting to calculate the complete NLO fragmentation function
and compare the result with their approximation.

\subsection{Gluon fragmentation}

If a Higgs boson has momentum smaller than $2 M_t$,
its couplings to gluons can be described by the vertices of an effective field theory 
in which the top quark has been integrated out.
There is a region of $P_T$ above the Higgs mass  
$M_H$ but not too far above the top-quark-pair threshold $2M_t+M_H$
in which gluon fragmentation into Higgs can be calculated using this effective theory.
The LO fragmentation function for gluon  
into Higgs  comes from the tree-level process $g^* \to H+g$
with an effective $ggH$ vertex.
The phase space integral is quadratically ultraviolet divergent.
In the $\overline{\text{MS}}$ factorization scheme,
the quadratic divergence is removed by analytic continuation,
and the fragmentation function  is order $\alpha_s^2 M_H^2/v^2$.
In the invariant-mass-cutoff scheme, the quadratic divergence gives an additional term 
proportional to $\alpha_s^2 \mu^2/v^2$.
We do not present this fragmentation function here,
because it is two orders higher in $\alpha_s$ than the fragmentation function for $t \to H$.

For $P_T$ well above the top-quark-pair threshold $2M_t+M_H$,
the leading-order contribution to the fragmentation function for gluon
into Higgs comes from the tree-level process $g^* \to H+ t \bar t$, and is order $\alpha_s y_t^2$.
It is smaller than the fragmentation function for $t \to H$ by a factor of $\alpha_s$.
We do not present this fragmentation function here,
because it is one order higher in $\alpha_s$ than the fragmentation function for $t \to H$.


\section{Evolution of Fragmentation Functions}
\label{sec:evolution}

If a fragmentation function for Higgs production is specified at some initial scale $\mu_0$,
 the solution of the evolution equation in Eq.~\eqref{eq:evo}
gives the fragmentation functions at a larger scale $\mu$,  
with the  leading logarithms of $\mu/\mu_0$ summed to all orders. 
In this section, we 
identify appropriate initial conditions for the fragmentation functions
in the $\overline{\text{MS}}$ scheme and in the IMC scheme. 
We calculate the effects of QCD evolution on the fragmentation function for $t \to H$.
We also calculate the fragmentation function for $g \to H$ induced by QCD evolution.

\subsection{Initial conditions}

In this work, we only consider the resummation of 
the leading logarithms from QCD  interactions, 
which are expected to be numerically dominant. 
At leading order in $\alpha_s$,
the evolution equation in Eq.~\eqref{eq:evo} reduces to
\begin{equation}
\label{eq:evoLO}
\mu^2\frac{\partial}{\partial \mu^2}D_{i\to H}(z,\mu) = P_{i\to H}(z,\mu)
+ \sum_{j\neq H} \int_z^1\frac{\td y}{y}P_{i\to j}(z/y,\mu)D_{j\to H}(y,\mu).
\end{equation}
The nonhomogenous term comes from the LO fragmentation function 
for $H \to H$ in Eq.~\eqref{eq:DHH}.
The homogeneous terms on the right side of Eq.~\eqref{eq:evoLO} are
nonzero only if the fragmenting parton $i$ is a quark or gluon.
The intermediate parton $j$ must also be a quark or gluon.
If the fragmentation function is known for all $z$ at an initial scale $\mu_0$,
the evolution equation in Eq.~\eqref{eq:evoLO} can be integrated
to obtain the fragmentation function at any higher scale $\mu$.

We first discuss the evolution of fragmentation functions in the invariant-mass-cutoff (IMC) scheme.
In Section~\ref{sec:FF}, we calculated
the LO fragmentation functions $D^{\text{IMC}}_{i\to H}(z,\mu)$
for $V \to H$ and $t\to H$ in this scheme.
The invariant mass constraint $\mu^2 > M_H^2/z + M_i^2/(1-z)$ 
for the tree-level process $i^* \to H+i$ implies that the LO fragmentation function is nonzero only 
for $z$ within a subinterval $(z_-,z_+)$ of the interval $(0,1)$.
As $\mu$ decreases,
the length $z_+-z_-$ of the subinterval decreases, 
 reaching 0 at the scale $\mu_{0,i} = M_H + M_i$.
It is natural to take  the vanishing of $D^{\text{IMC}}_{i\to H}(z,\mu)$ at this scale as the initial condition 
on the fragmentation function:
\begin{equation}
\label{eq:evomu0cut}
D^{\text{IMC}}_{i\to H}(z,\mu=\mu_{0,i}) = 0 .
\end{equation}
At larger scales $\mu$, $D^{\text{IMC}}_{i\to H}(z,\mu)$ is obtained by solving 
the inhomogeneous evolution equation in Eq.~\eqref{eq:evoLO}.
If we also define the LO fragmentation function to be 0 
for $ \mu <\mu_{0,i}$, it is a continuous function for all positive  $\mu$ and $0<z<1$.
To complete the prescription for the evolution of the fragmentation functions
in the IMC scheme,
we need to specify the splitting functions in Eq.~\eqref{eq:evoLO}.
We take the inhomogeneous term to be the splitting function
$P^{\text{IMC}}_{i\to H}(z,\mu)$  obtained by differentiating the 
LO fragmentation function in the IMC scheme, as in Eq.~\eqref{eq:PiH}.
We take the QCD splitting functions $P_{i\to j}(z,\mu)$ to be the 
LO splitting functions in the $\overline{\text{MS}}$ scheme.
The splitting function $P^{\text{IMC}}_{i\to H}(z,\mu)$ has a $\theta$ function factor
that ensures that the inhomogeneous term in the evolution equation in 
Eq.~\eqref{eq:evoLO} gives nonzero contributions only  for $z$ 
inside the subinterval $(z_-,z_+)$ in which the LO fragmentation function is nonzero.
However the homogeneous terms in Eq.~\eqref{eq:evoLO} will give 
nonzero contributions in the entire interval (0,1).

We next discuss the evolution of fragmentation functions in the $\overline{\text{MS}}$ scheme.
In Section~\ref{sec:FF}, we calculated the LO fragmentation functions 
$D_{i\to H}(z,\mu)$ for $V \to H$ and $t \to H$ in this scheme.
Since the renormalization scale $\mu$ in the $\overline{\text{MS}}$ scheme
has no direct physical interpretation, it is necessary to  
choose an initial scale $\mu_0$ for the fragmentation function.
A  simple choice for $\mu_0$ is the minimum invariant mass of the partons 
in the final state for the LO fragmentation process.
For the tree-level process $i^* \to H+i$, this scale is $\mu_{0,i} = M_H + M_i$.
Our initial condition on $D_{i \to H}(z,\mu)$ is equal to the LO fragmentation function at that scale: 
\begin{equation}
\label{eq:evomu0}
D_{i\to H}(z,\mu=\mu_{0,i}) = D^{\text{LO}}_{i\to H}(z,\mu_{0,i}) .
\end{equation}
At larger scales $\mu$, $D_{i\to H}(z,\mu)$ is obtained by solving 
the evolution equation in Eq.~\eqref{eq:evoLO}.
If we also define the LO fragmentation function to be 0 for $ \mu <\mu_{0,i}$, it changes discontinuously
from 0 for all $z$ to nonzero as $\mu$ increases through $\mu_{0,i}$.
To complete the prescription for the evolution of the fragmentation functions
in the $\overline{\text{MS}}$ scheme,
we need to specify the splitting functions in Eq.~\eqref{eq:evoLO}.
We take the inhomogeneous term to be the splitting function obtained by differentiating the LO fragmentation function 
in the $\overline{\text{MS}}$ scheme, as in Eq.~\eqref{eq:PiH}.
We take the QCD splitting functions $P_{i\to j}(z,\mu)$ to be the 
LO splitting functions in the $\overline{\text{MS}}$ scheme. 
Since the $\overline{\text{MS}}$ scheme ignores constraints 
on the invariant mass of final-state partons, the fragmentation function may not have a simple physical 
interpretation as a probability distribution for the longitudinal momentum 
fraction $z$ in a jet.  
In particular, it can be negative in some regions of $z$.
However, 
in the LP factorization formula in Eq.~\eqref{eq:LPfactorization}, 
the unphysical aspects of a fragmentation function are cancelled by 
a subtraction term that removes mass singularities from an infrared-safe cross section.
Thus the sum over fragmenting partons
gives a physical hard-scattering cross section $\td \hat \sigma$.

\subsection{Weak vector boson fragmentation}

Since we only resum the large logarithms due to  QCD interactions, 
the $W$ and $Z$ fragmentation functions 
decouple from the top-quark and gluon fragmentation functions.  
The only term in the LO evolution equation for the 
fragmentation function for a weak vector boson into Higgs
is the inhomogeneous term $P_{V \to H}$ in Eq.~\eqref{eq:evoLO}.
The solution to the evolution equation  for $\mu > \mu_{0,V}$
is therefore just the LO fragmentation function evaluated at the scale $\mu$.
The fragmentation functions for $W^\pm$ and $Z^0$ into Higgs at  
any scale $\mu$ are given by Eq.~\eqref{eq:V->H} in the $\overline{\text{MS}}$ scheme
and by Eq.~\eqref{eq:V->Hcutoff} in the IMC scheme.
The dependence of the fragmentation functions  
on $\mu$ are illustrated in Fig.~\ref{fig:V->H}.

\subsection{ QCD evolution equations at LO}

The QCD evolution equations for the top quark and gluon fragmentations are coupled
so they must be solved together. 
They have the schematic forms
\begin{subequations}
\label{eq:tgevo2}
\begin{eqnarray}
\mu^2\frac{\td}{\td \mu^2} D_{t\to H}
&=& P_{t\to H}
+ P_{t\to g} \otimes D_{g \to H} + P_{t\to t} \otimes D_{t\to H},
\label{eq:tevo2}
\\
\mu^2\frac{\td}{\td \mu^2} D_{g\to H}
&=& P_{g\to H}
+ P_{g\to g} \otimes D_{g\to H} + 2 P_{g\to t} \otimes D_{t\to H} .
\label{eq:gevo2}
\end{eqnarray}
\end{subequations}
The fragmentation functions from other quark flavors $q$ are not important, 
since they are suppressed either  by $\as$ or by $y_q^2/y_t^2$, 
where $y_q$ is the Yukawa coupling of quark flavor $q$.
The factor of $2$ in the last term in Eq.~\eqref{eq:gevo2} takes into account 
 the fragmentation of both $t$ and $\bar t$. 
The inhomogeneous term $P_{g\to H}$ in Eq.~\eqref{eq:gevo2} is order $\alpha_s^2 y_t^2$ for 
$\mu$ between $M_H$ and $2M_t+M_H$ and order $\alpha_s y_t^2$ for $\mu > 2M_t+M_H$.
The inhomogeneous term $P_{t\to H}$ in Eq.~\eqref{eq:tevo2} is 0 for $\mu < M_t+M_H$
and order $y_t^2$ for $\mu > M_t+M_H$.
The QCD splitting functions $P_{g\to g}$, $P_{g\to t}$, $P_{t\to g}$, and $P_{t\to t}$ are order $\alpha_s$.
The LO fragmentation functions $D_{i \to H}$ are the same order as $P_{i \to H}$.

The thresholds for $H$, $H+t$, and $H+t \bar t$ are the boundaries between three regions of $\mu$ 
in which the terms in the evolution equations that are LO are different:
 \begin{enumerate}
 \item
 $M_H < \mu < M_t+M_H$:
 In this region, the top quark can be integrated out completely and we can set $D_{t \to H}(z,\mu)=0$.
 The evolution equation for  $D_{g \to H}(z,\mu)$ reduces to  Eq.~\eqref{eq:gevo2}
 with the $P_{g\to H}$ and $P_{g \to g} \otimes D_{g \to H}$ terms only.
 \item
 $M_t+M_H < \mu < 2M_t+M_H$:
 In this region, we can set $D_{g \to H}(z,\mu)=0$,
 because $P_{g\to H}$ is suppressed by $\alpha_s$ compared to $P_{t\to H}$
 and because $\mu$ is below the threshold for the $P_{g\to t} \otimes D_{t\to H}$ term.
 The evolution equation for  $D_{t \to H}(z,\mu)$ reduces to  Eq.~\eqref{eq:tevo2}
 with the $P_{t\to H}$ and $P_{t \to t} \otimes D_{t \to H}$ terms only.
\item
 $\mu > 2M_t+M_H$:
In this region, we can set $P_{g\to H}$ to 0, because it is suppressed by $\alpha_s$ compared to $P_{t\to H}$.
All the other terms in  Eqs.~\eqref{eq:tgevo2} must be included.
\end{enumerate}

We consider the evolution of the fragmentation functions only at LO in $\alpha_s$, which is order $y_t^2$.
At this order, both the top-quark and the gluon fragmentation functions are zero in region 1.
In region 2, we impose the initial condition on the top-quark fragmentation function
at the scale $\mu_{0,t} = M_t+M_H$. In the $\overline{\text{MS}}$ scheme, 
$D_{t\to H}(z,\mu_{0,t})$ is obtained by setting $\mu = M_t+M_H$ in Eq.~\eqref{eq:t->H}.
In the IMC scheme, $D_{t\to H}(z,\mu_{0,t})$ is 0 for all $z$.
The fragmentation functions at  $M_t+M_H<\mu < 2M_t+M_H$ are obtained by solving the evolution equation
\begin{eqnarray}
\label{eq:tevol-1}
\mu^2\frac{\td}{\td \mu^2} D_{t\to H}(z,\mu)
&=& P_{t\to H}(z,\mu)
+ \int_z^1\frac{\td y}{y}P_{t\to t}(z/y,\mu) D_{t\to H}(y,\mu).
\end{eqnarray}
In the $\overline{\text{MS}}$ scheme, the $t \to H$ splitting function $P_{t\to H}(z,\mu)$
is independent of $\mu$ and is given in Eq.~\eqref{P:t->H}.
In the IMC scheme, the $t \to H$ splitting function is given in Eq.~\eqref{eq:Pt->Hcut}.
In this region, the gluon fragmentation function is identically zero in both $\overline{\text{MS}}$ and IMC schemes.

In region 3, we impose as the initial condition on the top quark fragmentation function
at the scale $\mu_{0,t\bar t} = 2M_t+M_H$ the result obtained from integrating Eq.~\eqref{eq:tevol-1}. 
The initial gluon fragmentation function $D_{g\to H}(z,\mu_{0,t \bar t})$ is 0 for all $z$.
The fragmentation functions at larger $\mu$ are obtained by solving the evolution equations
\begin{subequations}
\label{eq:tgevol}
\begin{eqnarray}
\label{eq:tevol}
\mu^2\frac{\td}{\td \mu^2} D_{t\to H}(z,\mu)
&=& P_{t\to H}(z,\mu)
+ \int_z^1\frac{\td y}{y}P_{t\to t}(z/y,\mu) D_{t\to H}(y,\mu)
\nonumber\\
&&
+ \int_z^1\frac{\td y}{y}P_{t\to g}(z/y,\mu) D_{g\to H}(y,\mu),
\\
\mu^2\frac{\td}{\td \mu^2} D_{g\to H}(z,\mu)
&=&  \int_z^1\frac{\td y}{y}P_{g\to g}(z/y,\mu) D_{g\to H}(y,\mu)
+ 2 \int_z^1\frac{\td y}{y} P_{g\to t}(z/y,\mu) D_{t\to H}(y,\mu).
\nonumber\\
\label{eq:gevol}
\end{eqnarray}
\end{subequations}

\subsection{Top-quark fragmentation}

\begin{figure}
\begin{center}
\includegraphics[width=0.82\textwidth]{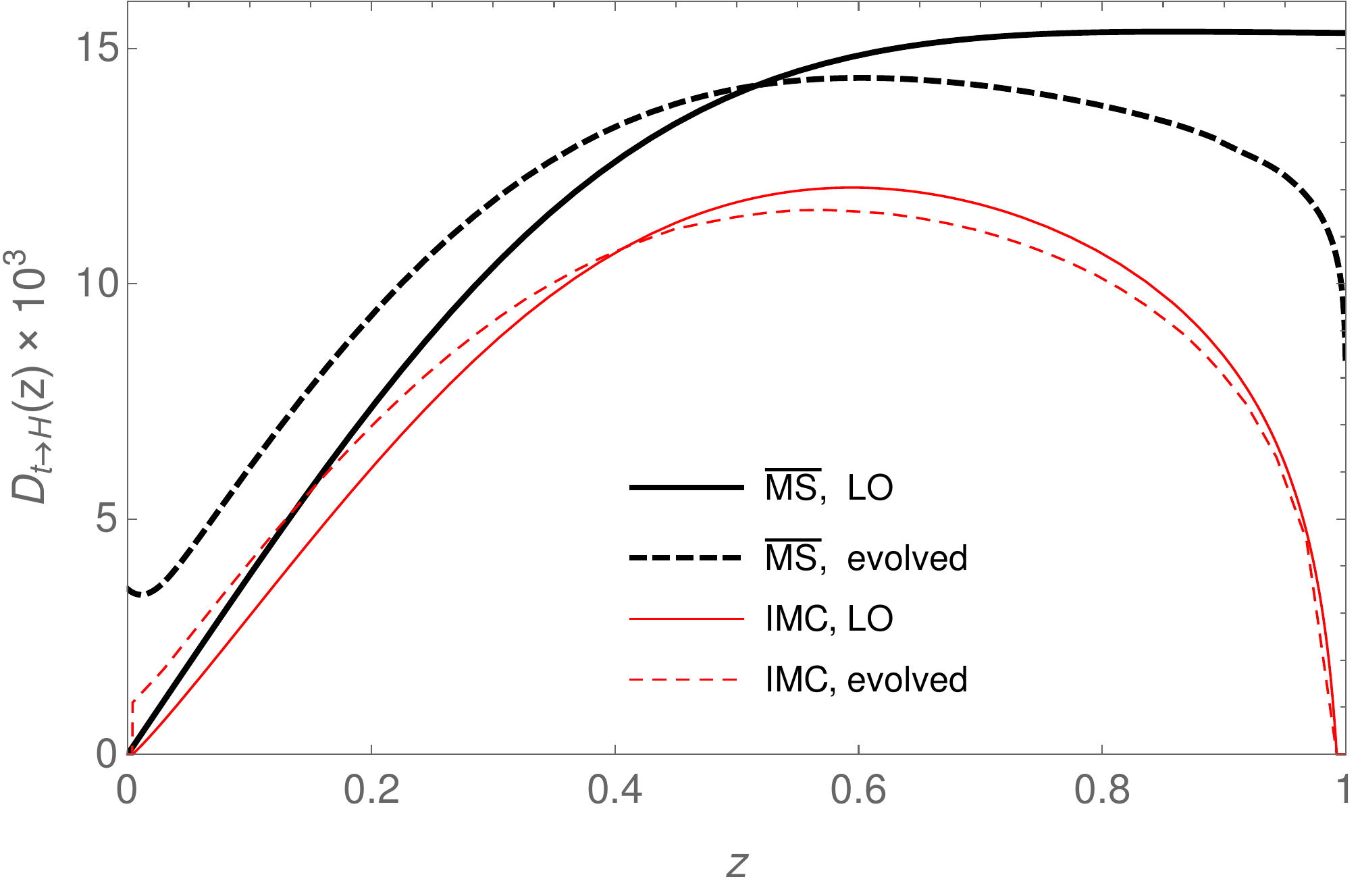}
\caption{LO fragmentation functions (solid curves)
and evolved fragmentation functions (dashed curves) for top quark into Higgs 
at $\mu = 2$~TeV 
in the $\overline{\text{MS}}$ scheme (thicker curves) 
and in the IMC scheme (thinner curves).
\label{fig:t->Hevo}}
\end{center}
\end{figure}

The effect of evolution on the fragmentation function for 
top quark into Higgs is illustrated in Fig.~\ref{fig:t->Hevo}
by showing fragmentation functions at the scale $\mu=2$~TeV.
The LO fragmentation function 
in the $\overline{\text{MS}}$ scheme at that scale
is obtained by setting $\mu = 2$~TeV in Eq.~\eqref{eq:t->H}. 
The LO fragmentation function 
in the IMC scheme at that scale 
is obtained by setting $\mu = 2$~TeV in Eq.~\eqref{eq:t->Hcut}. 
The corresponding evolved fragmentation functions at that scale
are obtained by integrating the coupled differential equations in
Eqs.~\eqref{eq:tgevol} from the initial scale up to 2~TeV.

In both the $\overline{\text{MS}}$ and IMC schemes,  
the effect of evolution on the top-quark fragmentation function is to suppress it at large $z$ 
and to enhance it at small $z$. 
The effect of evolution is milder in the IMC scheme 
compared to the $\overline{\text{MS}}$ scheme.
Because of the rapid decrease in the parton distributions of the proton at large $x$, 
the production of Higgs in $pp$ collisions is dominated by fragmentation at large $z$. 
Consequently, the evolution of the top-quark fragmentation 
suppresses the production of the Higgs collinear to a top quark.

\subsection{Gluon fragmentation}

\begin{figure}
\begin{center}
\includegraphics[width=0.8\textwidth]{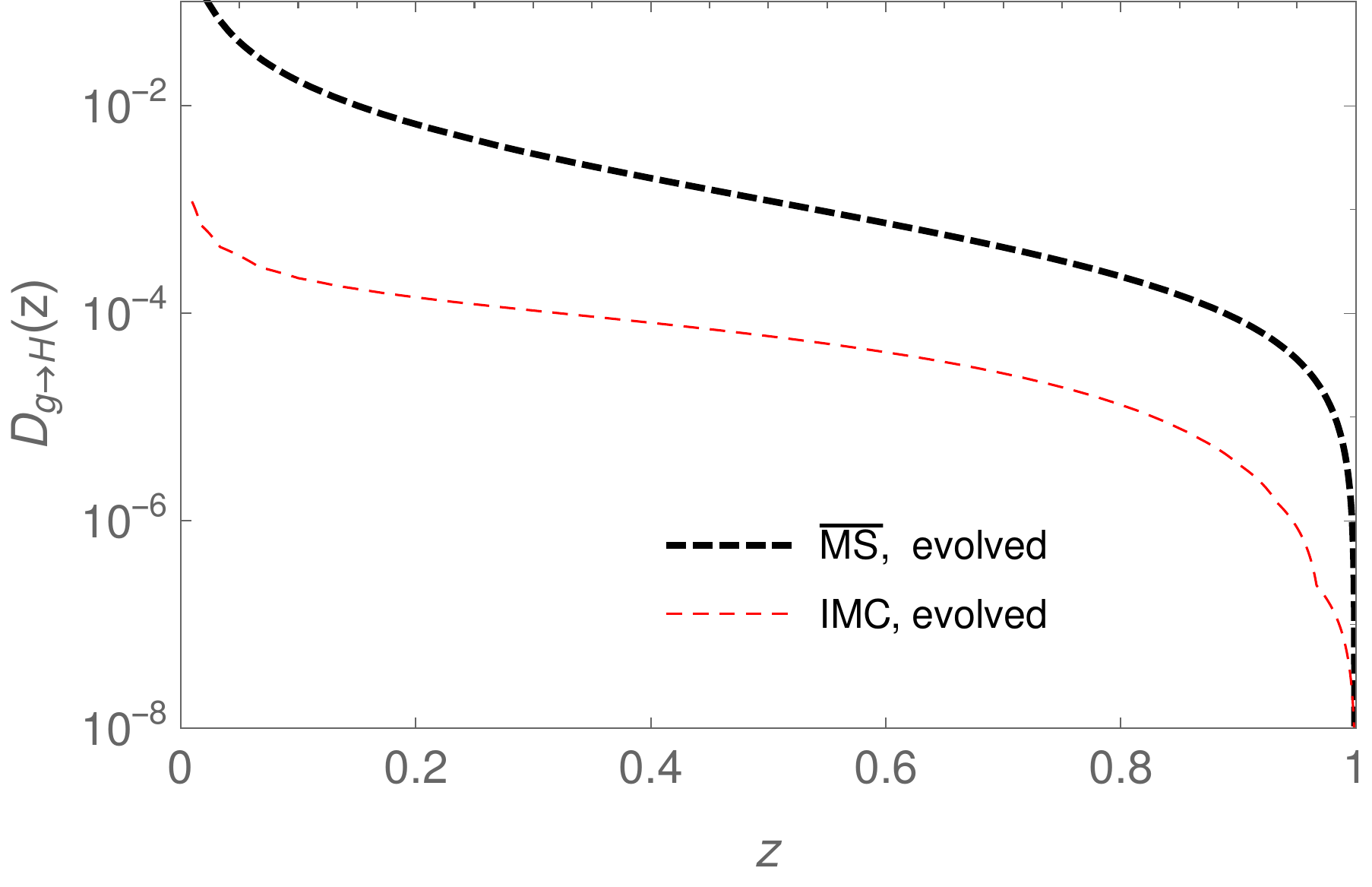}
\caption{ 
Evolved fragmentation functions  for gluon into Higgs
at $\mu = 2$~TeV
in the $\overline{\text{MS}}$ scheme (thicker curve) 
and in the IMC scheme (thinner curve).
\label{fig:g->Hhi}}
\end{center}
\end{figure}

The effect of evolution on the fragmentation function for gluon into Higgs  is illustrated in Fig.~\ref{fig:g->Hhi}
by showing fragmentation functions at the scale $\mu=2$~TeV.
The fragmentation function at order $y_t^2$ is zero.
However the evolved fragmentation functions
obtained by solving the coupled differential equations in
Eqs.~\eqref{eq:tgevol} with the appropriate initial conditions
have terms of order $y_t^2[\as \log(\mu^2/M_H^2)]^n$, which is order $y_t^2$ for sufficiently large $\mu$.
The evolved fragmentation functions in the $\overline{\text{MS}}$ scheme 
and  in the IMC scheme are shown at the scale 2~TeV.

In both the $\overline{\text{MS}}$ and IMC schemes, 
the initial gluon fragmentation function is zero at $\mu=2M_t+M_H$. 
The nonzero value is initially generated by the second term on the right side of Eq.~\eqref{eq:gevol}. 
 The gluon fragmentation function  grows more rapidly
at smaller $z$ due to the larger integration region of $y$ on the right side of Eq.~\eqref{eq:gevol}. 
 The gluon fragmentation function is  much smaller
in the IMC scheme compared to $\overline{\text{MS}}$ scheme.
At $\mu=2$~TeV, the gluon fragmentation function at small $z$ is almost as large as 
the top-quark fragmentation function. However, the gluon fragmentation function at large $z$ region 
is only approximately $1\%$ of the top-quark fragmentation function. 
Since the production of Higgs in $pp$ collisions is dominated by fragmentation at large $z$,
gluon fragmentation can be ignored compared with top-quark fragmentation.


\section{Comparison with a complete LO calculation}\label{sec:compare}

In this section, we apply the LP factorization formula in Eq.~\eqref{eq:LPfactorization} 
to the process $q\bar{q}\to H t\bar{t}$ at LO, where $q$ is a massless quark. 
We compare the LO cross sections from the LP factorization formula  using  the zero-mass-top-quark 
and hybrid factorization prescriptions 
to the complete LO cross section.
We estimate the minimum Higgs transverse momentum $P_T$ 
above which the LP factorization formula is reliable.

\subsection{ Infrared-safe cross sections}

\begin{figure}
\begin{center}
\includegraphics[width=0.3\textwidth]{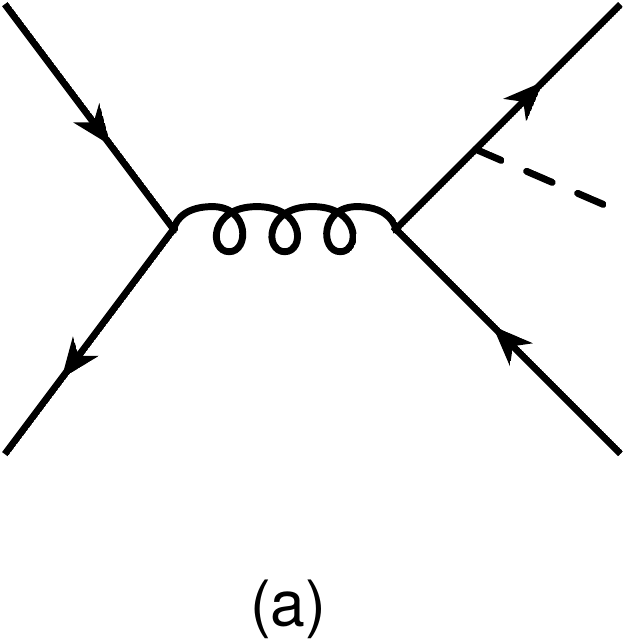}
\hspace{2cm}
\includegraphics[width=0.3\textwidth]{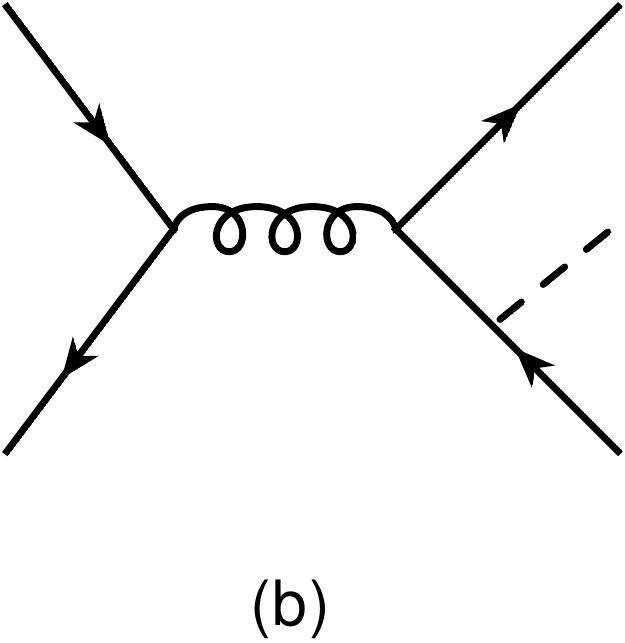}
\caption{Feynman diagrams for the hard-scattering cross section 
$\hat{\sigma}_{q\bar{q}\to H+t\bar t}$ at leading order in the coupling constants.
\label{fig:FG}}
\end{center}
\end{figure}

The process $q\bar{q}\to H t\bar{t}$ proceeds at LO through the Feynman diagrams 
in Figure~\ref{fig:FG}.  The LO cross section is order $\alpha_s^2 y_t^2$.
The complete expression for the differential cross section $\td\hat{\sigma}/\td P_T^2 \td \hat y$,
where $P_T$ is the transverse momentum of the Higgs and 
$\hat y$ is its rapidity in the $q \bar q$ center-of-momentum frame,
is given in Appendix~\ref{app:LO}.
It has a mass singularity in  the limits $M_H\to 0$ and $M_t\to 0$.

At LO, the LP factorization formula in Eq.~\eqref{eq:LPfactorization} 
for the process $q\bar{q}\to H t\bar{t}$ has two terms:
\begin{equation}
\label{eq:LPfacformula_LO}
\td\hat{\sigma}_{q\bar{q}\rightarrow H t\bar t }^{\textrm{(fp)}}(P)
= 
\td\tilde{\sigma}_{q\bar{q}\rightarrow H+t\bar t}^{\textrm{(fp)}}(\tilde{P},\mu)
+2\int_0^1 \td z\, 
\td\tilde{\sigma}_{q\bar{q}\rightarrow t+\bar{t}}^{\textrm{(fp)}}
(p=\tilde{P}/z) \,
D_{t\rightarrow H}(z,\mu),
\end{equation}
where $P$ is the momentum of the Higgs with mass $M_H$, 
$\tilde{P}$ is the corresponding momentum for a massless Higgs
defined in Eq.~\eqref{eq:tildeP-covaraint}, and
$p$ is the momentum of the fragmenting top quark.
We call the two terms on the RHS of 
Eq.~\eqref{eq:LPfacformula_LO} the direct contribution
and the fragmentation contribution.
We have explicitly used the fact that the fragmentation contributions 
from $t$ and $\bar{t}$ are the same.
The superscripts (fp) in Eq.~\eqref{eq:LPfacformula_LO} indicate the terms that depend at this order 
on the factorization prescription.
The factorization prescriptions we consider are the massive-top-quark (MTQ) prescription,
the zero-mass-top-quark (ZMTQ) prescription, and the hybrid prescription,
which were described in Section~\ref{sec:topmass}.

The infrared-safe cross sections $\td\tilde{\sigma}$ in Eq.~\eqref{eq:LPfacformula_LO}
are constructed order-by-order in the coupling constants
from the hard-scattering cross sections $\td\hat{\sigma}$.
The infrared-safe cross section for $q\bar{q}\rightarrow H+t\bar t$ is order $\alpha_s^2 y_t^2$.
The infrared-safe cross section for $q\bar{q}\rightarrow t + \bar t$ is order $\alpha_s^2$.
The fragmentation function for $t \to H$ is order $y_t^2$.
We begin at order $\alpha_s^2$ by determining the infrared-safe cross section  $\td\tilde{\sigma}$
for $q\bar{q}\rightarrow t + \bar t$.  This is easy because the corresponding
hard-scattering cross section $\td\hat{\sigma}$ is already infrared safe.
Depending on the factorization prescription, the limit $M_t \to 0$ may or may not be taken
in this cross section.
Having determined $\td\tilde{\sigma}$ for $q\bar{q}\rightarrow t + \bar t$,
we proceed to the next order in the coupling constants, which is $\alpha_s^2 y_t^2$.
After replacing $\td\hat{\sigma}$ on the left side of Eq.~\eqref{eq:LPfacformula_LO}
by the complete LO hard-scattering cross section for $q\bar{q}\rightarrow H t\bar t$, 
we solve for the infrared-safe cross section for $q\bar{q}\rightarrow H + t\bar t$:
\begin{equation}
\label{eq:tildesigma-Htt_LO}
\td\tilde{\sigma}_{q\bar{q}\rightarrow H+t\bar t}^{\textrm{(fp)}}(\tilde{P},\mu) =
\td\hat{\sigma}_{q\bar{q}\rightarrow H t\bar t }(P)
- 2\int_0^1 \td z\, 
\td\tilde{\sigma}_{q\bar{q}\rightarrow t+\bar{t}}^{\textrm{(fp)}}
(p=\tilde{P}/z) \,
D_{t\rightarrow H}(z,\mu).
\end{equation}
The limit $M_H \to 0$ is taken on the right side.
Depending on the factorization prescription, the limit $M_t \to 0$ may or may not also be taken.

In the MTQ prescription, $M_t$ is kept at its physical value in both 
infrared-safe cross sections:
\begin{subequations}
\begin{eqnarray}
\label{eq:IRScc1-mqt}
\td\tilde{\sigma}_{q\bar{q}\rightarrow t + \bar t }^{\textrm{(MTQ)}}(p) &=&
\td\hat{\sigma}_{q\bar{q}\rightarrow t \bar{t}}(p).
\\
\label{eq:IRScc2-mqt}
\td \tilde{\sigma}_{q\bar{q}\rightarrow H+t\bar t }^{\textrm{(MTQ)}}(P,\mu) &=& 
 \lim_{M_H \to 0} \td \hat{\sigma}_{q\bar{q}\rightarrow Ht\bar t}(P)
- 2 \int_0^{z_{\textrm{max}}}\!\!\!\!\!\! \td z \, 
\td \tilde{\sigma}_{q\bar{q}\rightarrow t+\bar{t}}^{\textrm{(MTQ)}}
(p \approx\tilde{P}/z) \,
 \lim_{M_H \to 0} D_{t\rightarrow H}(z,\mu),
\nonumber\\
\end{eqnarray}
\label{eq:IRScc-mqt2}%
\end{subequations}
where $p\approx\tilde{P}/z$ represents the complicated expression 
for the top-quark 4-momentum in Eq.~\eqref{eq:pt} that respects the mass-shell constraint $p^2 = M_t^2$
and a longitudinal momentum constraint.
The upper limit $z_{\textrm{max}}$ on the integral over $z$ is given in Eq.~\eqref{eq:zmax}.
The limit of the fragmentation function as $M_H \to 0$ is obtained 
by setting $\zeta_t = 0$ in Eq.~\eqref{eq:t->H} or \eqref{eq:t->Hcut}.
The subtraction on the right side of Eq.~\eqref{eq:IRScc2-mqt}
cancels the cross section in the collinear region, in which the Higgs is produced collinearly with the top quark. 
Both terms on the right side of Eq.~\eqref{eq:IRScc2-mqt}
have mass singularities in the limit $M_t \to 0$,
but those singularities are regularized by the top-quark mass. 
The simplifications provided by the LP factorization formula compared with the complete LO calculation 
are not evident from the presentation above, because we used 
the hard-scattering cross section $\td \hat{\sigma}$ for $q \bar q \to Ht \bar t$ that depends on $M_t$ and $M_H$. However we could have
set $M_H=0$ from the beginning in the calculation of this cross section,
which would have made the calculation much simpler. 
This introduces fractional errors of order $M_H^2/M_t^2$ in the collinear region 
and fractional errors of order $M_H^2/P_T^2$ in the non-collinear region. 
Since the collinear region is subtracted in Eq.~\eqref{eq:IRScc-mqt2}, 
the fractional error in the LP factorization formula in Eq.~\eqref{eq:LPfacformula_LO} 
is order $M_H^2/P_T^2$.

In the ZMTQ prescription, the limit $M_t \to 0$ is taken in both infrared-safe cross sections:
\begin{subequations}
\begin{eqnarray}
\label{eq:IRScc1-zmqt}
\td\tilde{\sigma}_{q\bar{q}\rightarrow t + \bar t }^{\textrm{(ZMTQ)}}(p) &=&
\lim_{M_t \to 0} \td\hat{\sigma}_{q\bar{q}\rightarrow t \bar{t}}(p),
\\
\label{eq:IRScc2-zmqt}
\td \tilde{\sigma}_{q\bar{q}\rightarrow H+t\bar t }^{\textrm{(ZMTQ)}}(\tilde{P},\mu) &=& 
 \lim_{M_t \to 0}  \Bigg(  \lim_{M_H \to 0} \td \hat{\sigma}_{q\bar{q}\rightarrow Ht\bar t}(P)
 \nonumber\\
&& \hspace{1.5cm}
- 2\int_0^1\!\!\! \td z \, 
\td \tilde{\sigma}_{q\bar{q}\rightarrow t+\bar{t}}^{\textrm{(ZMTQ)}}
(p=\tilde{P}/z) \,
 \lim_{M_H \to 0} D_{t\rightarrow H}(z,\mu) \Bigg),
\end{eqnarray}
\label{eq:IRScc-zmqt}%
\end{subequations}
where $\tilde P$ is the light-like momentum defined in Eq.~\eqref{eq:tildeP-covaraint}.
The subtracted term on the right side of Eq.~\eqref{eq:IRScc2-zmqt}
cancels the mass singularity in the first term, 
leaving a cross section that  has a well-behaved limit as $M_t\to 0$. 
The fractional error in the LP factorization formula is of order $M_t^2/P_T^2$.
The calculation of the infrared-safe cross section for $q \bar q \to H+t \bar t$
can be greatly simplified by setting $M_H=0$ from the beginning in the calculation of 
the hard-scattering cross section $\td \hat{\sigma}$ for $q \bar q \to H t \bar t$.
The calculation can be simplified much further by setting both $M_H=0$  and $M_t=0$ 
from the beginning in the calculation of the hard-scattering cross sections $\td \hat{\sigma}$ 
for $q \bar q \to Ht \bar t$ and $q \bar q \to t \bar t$,
and by using dimensional regularization in $D$ spacetime dimensions to regularize the mass singularities. 
The mass singularities cancel between the two terms on the right side of Eq.~\eqref{eq:IRScc2-zmqt},
and the same infrared-safe cross section $\td \tilde{\sigma}$ is obtained in the limit $D \to 4$.
With the ZMTQ prescription, 
the absence of any mass scales in the infrared-safe cross sections 
dramatically simplifies the calculation of higher order corrections.

In the hybrid prescription, 
the physical value of $M_t$ is used in the infrared-safe cross section for $q \bar q \to H + t \bar t$, 
but the limit $M_t \to 0$ is taken in the infrared-safe cross section for $q \bar q \to t +\bar t$ 
because the top quark is the fragmenting parton:
\begin{subequations}
\begin{eqnarray}
\label{eq:IRScc1-hybrid}
\td\tilde{\sigma}_{q\bar{q}\rightarrow t + \bar t }^{\textrm{(hybrid)}}(p) &=&
\lim_{M_t \to 0} \td\hat{\sigma}_{q\bar{q}\rightarrow t \bar{t}}(p),
\\
\label{eq:IRScc2-hybrid}
\td \tilde{\sigma}_{q\bar{q}\rightarrow H+t\bar t }^{\textrm{(hybrid)}}(\tilde{P},\mu) &=& 
\lim_{M_H \to 0} \td \hat{\sigma}_{q\bar{q}\rightarrow Ht\bar t}(P)
- 2 \int_0^1\!\!\! \td z \, 
\td \tilde{\sigma}_{q\bar{q}\rightarrow t+\bar{t}}^{\textrm{(hybrid)}}
(p=\tilde{P}/z) \,
 \lim_{M_H \to 0} D_{t\rightarrow H}(z,\mu).
 \nonumber\\
\end{eqnarray}
\label{eq:IRScc-hybrid}%
\end{subequations}
The error with the hybrid description is order $M_H^2/P_T^2$,
the same as with the MTQ prescription, which may be somewhat surprising.
This can be seen explicitly at LO by subtracting the LP factorization formulas 
with the two prescriptions, and expressing the  infrared-safe cross sections
 $\td \tilde \sigma$ in terms of hard-scattering cross sections  $\td \hat \sigma$
 and their limits as $M_t\to 0$ and $M_H \to 0$.
The cross sections $\td \hat{\sigma}$ for $q \bar q \to H t \bar t$ cancel,
and the remaining terms reduce to
\begin{eqnarray}
\label{eq:hybrid-mtq}
&&\td \hat{\sigma}_{q\bar{q}\rightarrow Ht\bar t }^{\textrm{(hybrid)}}(\tilde P,\mu) 
- \td \hat{\sigma}_{q\bar{q}\rightarrow Ht\bar t }^{\textrm{(MTQ)}}(P,\mu) 
 \nonumber\\
&&
= 2 \int_0^1\!\! \td z \, 
\left( \lim_{M_t \to 0} \td \hat{\sigma}_{q\bar{q}\rightarrow t\bar{t}} (p=\tilde{P}/z) \right)
\left[ D_{t\rightarrow H}(z,\mu)
- \lim_{M_H \to 0} D_{t\rightarrow H}(z,\mu) \right]
 \nonumber\\
&& \hspace{0.5cm}
 - 2 \int_0^{z_{\textrm{max}}}\!\! \td z \, 
\td \hat{\sigma}_{q\bar{q}\rightarrow t\bar{t}} (p\approx\tilde{P}/z)
 \left[ D_{t\rightarrow H}(z,\mu)
- \lim_{M_H \to 0} D_{t\rightarrow H}(z,\mu) \right].
\end{eqnarray}
The difference between the fragmentation functions is order $M_H^2/M_t^2$.
The difference between the hard-scattering cross sections for $q\bar{q}\rightarrow t\bar{t}$ 
is order $M_t^2/P_T^2$.
Thus the difference between the hard-scattering cross sections 
for $q\bar{q}\rightarrow Ht\bar t$ is order $M_H^2/P_T^2$.

There are large differences between the LO fragmentation functions in the  $\overline{\textrm{MS}}$
scheme and the IMC scheme that were calculated in Section~\ref{sec:FF}.
Despite those large differences, the difference between the LP factorization formulas 
in the two schemes is small.
The cross sections $\td \hat{\sigma}$ for $q \bar q \to H t \bar t$ cancel,
and the remaining terms reduce to
\begin{eqnarray}
\label{eq:MSbar-IMC}
&&\td \hat{\sigma}_{q\bar{q}\rightarrow Ht\bar t }^{(\textrm{fp},\overline{\textrm{MS}})}(\tilde P,\mu) 
- \td \hat{\sigma}_{q\bar{q}\rightarrow Ht\bar t }^{\textrm{(fp,IMC)}}(P,\mu) 
 \nonumber\\
&&
= 2 \int_0^1\!\! \td z \, 
\td \tilde{\sigma}_{q\bar{q}\rightarrow t\bar{t}}^{\textrm{(fp)}} (p=\tilde{P}/z) 
\left[ D_{t\rightarrow H}(z,\mu)
- \lim_{M_H \to 0} D_{t\rightarrow H}(z,\mu) \right]
 \nonumber\\
&& \hspace{0.5cm}
 - 2 \int_{z_-}^{z_+}\!\! \td z \, 
\td \tilde{\sigma}_{q\bar{q}\rightarrow t\bar{t}}^{\textrm{(fp)}} (p=\tilde{P}/z)
 \left[ D_{t\rightarrow H}^{\textrm{IMC}}(z,\mu)
- \lim_{M_H \to 0} D_{t\rightarrow H}^{\textrm{IMC}}(z,\mu) \right],
\end{eqnarray}
where $z_-$ and $z_+$ are given in Eq.~\eqref{eq:z+-}.
Inside the interval $(z_-,z_+)$, the difference between the fragmentation functions is order 
$M_t^2/\mu^2$ and the subtractions of their $M_H \to 0$ limits reduces the difference further
 to order $M_H^2/\mu^2$.  If $ \mu$ is chosen to be order $P_T$,
the contribution to the difference in the hard-scattering cross sections is order $M_H^2/P_T^2$.  
Since $z_- \approx M_H^2/\mu^2$ and $1-z_+ \approx M_t^2/\mu^2$ for $\mu \gg M_t$ and 
since the subtraction of the $M_H\to 0$ limit of the fragmentation function
makes the integrand of order $M_H^2/M_t^2$,
the contributions from the endpoint regions are also suppressed by at least $M_H^2/\mu^2$. 
This is order $M_H^2/P_T^2$ if $\mu$ is chosen to be of order $P_T$.

In the following subsections, we focus on the double differential cross section 
as a function of the Higgs transverse momentum $P_T$ and its rapidity $\hat{y}$ 
in the $q\bar q$ center-of-momentum frame. 
We consider only the hybrid prescription and the ZMTQ prescription. 
With these prescriptions, the top quark is treated as massless in the 
infrared-safe cross section for $q \bar q \to t+ \bar t$,
so its rapidity coincides with the rapidity $\hat{y}$ of the Higgs.
The LP factorization formula at LO is given by Eq.~\eqref{eq:LPfacformula_LO}:
\begin{equation}
\label{eq:fac_formula_LO_PT}
\frac{\td^2\hat{\sigma}_{q\bar{q}\rightarrow H t\bar t }^{\textrm{(fp)}}}{\td P_T^2 \td \hat y}
=
\frac{\td^2\tilde{\sigma}_{q\bar{q}\rightarrow H+t\bar t}^{\textrm{(fp)}}}{\td P_T ^2\td \hat y}
+ 2\int_0^1 \frac{\td z}{z^2}\, 
\frac{\td^2\tilde{\sigma}_{q\bar{q}\rightarrow t+\bar{t}}^{\textrm{(fp)}}}{\td p_T^2 \td \hat y}
(p=\tilde{P}/z) \,
D_{t\rightarrow H}(z).
\end{equation}
The infrared-safe differential cross-sections 
in the hybrid prescription are  given by  Eq.~\eqref{eq:IRScc-hybrid}: 
\begin{subequations}
\label{eq:SDCPT}
\begin{eqnarray}\label{eq:SDC1}
\frac{\td^2\tilde{\sigma}_{q\bar{q}\rightarrow t +\bar  t }^{\textrm{(hybrid)}}}{\td p_T^2 \td \hat y}
&=& 
\lim_{M_t\to 0}\frac{\td^2\hat{\sigma}_{q\bar{q}\rightarrow t \bar{t}}}{\td p_T^2 \td \hat y},
\\
\label{eq:SDC2}
\frac{\td^2\tilde{\sigma}_{q\bar{q}\rightarrow H+t\bar t }^{\textrm{(hybrid)}}}{\td P_T^2 \td \hat y}
&=& 
\lim_{M_H\to 0}
\frac{\td^2\hat{\sigma}_{q\bar{q}\rightarrow H t\bar t}}{\td P_T^2 \td \hat y}
-2\int_0^1 \frac{\td z}{z^2}\, 
\frac{\td^2\tilde{\sigma}_{q\bar{q}\rightarrow t+\bar{t}}^{\textrm{(hybrid)}}}{\td p_T^2 \td \hat y}(p=\tilde{P}/z)
\lim_{M_H\to 0}D_{t\rightarrow H}(z) .
\end{eqnarray}
\end{subequations}
In the ZMTQ prescription, the limit $M_t \to 0$ is also taken in Eq.~\eqref{eq:SDC2}.
The calculations of these two infrared-safe cross sections are presented in the next two subsections.

\subsection{Fragmentation contribution}

The fragmentation contribution to the LO cross section for $q\bar{q}\to H t\bar{t}$
is the second term on the right side of   Eq.~\eqref{eq:fac_formula_LO_PT}.
The fragmentation function $D_{t\to H}(z,\mu)$ is calculated in the $\overline{\text{MS}}$ factorization scheme
in Eq.~\eqref{eq:t->H}  and in the IMC  factorization scheme in Eq.~\eqref{eq:t->Hcut}.
To complete the calculation of the fragmentation contribution to the LP factorization formula in 
 Eq.~\eqref{eq:fac_formula_LO_PT}, we
 only need to calculate the infrared-safe cross section for $q\bar{q}\rightarrow t +\bar{t}$. 
We denote the momenta of $q$, $\bar q$ and the fragmenting top quark
by $k_1$, $k_2$ and $p$, respectively. 
The Mandelstam variables are $\hat{s}=(k_1+k_2)^2$, $\hat{t}=(k_1-p)^2$ and $\hat{u}=(k_2-p)^2$.
In both the  hybrid prescription and the ZMTQ prescription, we set $M_t=0$ in the cross section.  
The corresponding averaged matrix element at LO is
\begin{equation}
\overline{|\mathcal{M}|^2}_{q\bar{q}\to t\bar{t}}
=\frac{4\,(4\pi)^2\as^2}{9\,\hat{s}^2}\left(\hat{t}\,^2+\hat{u}^2\right),
\end{equation}
In the double differential cross section in the transverse momentum $p_T$ 
and rapidity $\hat y$ of the top quark in the $q \bar q$ center-of-momentum frame,
the integral over the $t +\bar{t}$ phase space is over-constrained.
We convert the extra $\delta$-function into a $\delta$-function of $z$ using $p=\tilde{P}/z$: 
\begin{equation}
\label{eq:SDCt}
\frac{\td^2\hat{\sigma}_{q\bar{q}\rightarrow t \bar{t}}}{\td p_{T}^2\,\td \hat y}(p=\tilde{P}/z) =
\frac{2\,\pi \,\as^2 \, z}{9\,\hat{s}^{\,2}}
\left(1+\tanh^2\hat y\right)\delta\Big(z- 2\cosh \hat y\, P_T /\sqrt{\hat s}\Big),
\end{equation}
where $P_T$ and $\hat y$ are the transverse momentum and the rapidity of the Higgs boson 
in the $q \bar q$ center-of-momentum frame. 
The integral over  $z$ in Eq.~\eqref{eq:fac_formula_LO_PT} is then trivial.
 Inserting the $\overline{\text{MS}}$ fragmentation function in Eq.~\eqref{eq:t->H} into  Eq.~\eqref{eq:fac_formula_LO_PT}
and integrating over $z$, we obtain the fragmentation contribution
\begin{eqnarray}
\label{eq:FragCont}
\int_0^1 \frac{\td z}{z^2}\, 
\frac{\td^2\tilde{\sigma}_{q\bar{q}\rightarrow t+\bar{t}}}{\td p_T^2 \td \hat y}
(p=\tilde{P}/z) \,
D_{t\rightarrow H}(z,\mu) &=&
\frac{\as^2 \,y_t^2}{72\,\pi\hat{s}^{\,2}}(1+\tanh^2 \hat{y})
\nonumber\\
&& \hspace{-2cm} \times
\bigg\{\log\frac{\mu^2}{4M_t^2}-\log\frac{\cosh^2 \hat{y}\, P_T^2+\zeta_t\,\sqrt{\hat s}\,(\sqrt{\hat s}-2\cosh\hat{y}\,P_T)\,}{\hat s}
\nonumber\\
&& \hspace{-1cm} +
\frac{(1-\zeta_t)\sqrt{\hat{s}}\,(\sqrt{\hat{s}}-2 \cosh \hat y\, P_T)} 
       {\cosh^2 \hat y\, P_T^2+ \zeta_t\,\sqrt{\hat s}\,(\sqrt{\hat s}-2\cosh \hat y\, P_T )}
\bigg\}.
\end{eqnarray}
The logarithm of $1/M_t^2$ becomes a mass singularity in the limit $M_t \to 0$.

\subsection{Direct contribution}

The direct contribution to the LO cross section for $q\bar{q}\to H t\bar{t}$
is the first term on the right side of Eq.~\eqref{eq:fac_formula_LO_PT}.
The infrared-safe cross section is defined in Eq.~\eqref{eq:SDC2}.
The subtraction term can be calculated in Eq.~\eqref{eq:FragCont} by setting $\zeta_t\to 0$. 
The other term is the differential cross section  
$\td^2\hat{\sigma}_{q\bar{q}\to Ht\bar t}/\td P_T^2 \td \hat y$,
which can be calculated from the
two Feynman diagrams  in Fig.~\ref{fig:FG}. 
The averaged matrix element can be expressed as
\begin{equation}\label{eq:ME}
\overline{|\mathcal{M}|^2} =
\overline{|\mathcal{M}|_{\text{aa}}^2} + \overline{|\mathcal{M}|_{\text{ab}}^2}
+ \overline{|\mathcal{M}|_{\text{ba}}^2} + \overline{|\mathcal{M}|_{\text{bb}}^2},
\end{equation}
where  $\overline{|\mathcal{M}|_{\text{ij}}^2}$
is the product of the amplitude for Feynman diagram $i$ 
and the complex conjugate of the amplitude for Feynman diagram $j$, 
averaged over initial spins and colors and summed over final spins and colors.
We can express the differential cross section as
\begin{equation}
\label{eq:Httbar0}
\frac{\td^2\hat{\sigma}_{q\bar{q}\rightarrow Ht\bar t }}{\td P_T^2 \td \hat y}
=
2\left(\frac{\td^2\hat{\sigma}^{\text{aa} }}{\td P_T^2 \td \hat y}
+\frac{\td^2\hat{\sigma}^{\text{ab} }}{\td P_T^2 \td \hat y}\right),
\end{equation}
where the superscripts have the same meaning as the subscripts in Eq.~\eqref{eq:ME}.

The 3-body phase space integral can be expressed as an iterated integral 
over the phase space of the Higgs and the 2-body phase space of $t\bar t$.
We denote the momenta of  $q$, $\bar q$, and the Higgs by $k_1$, $k_2$ and $P$, respectively.
A convenient set of Lorentz invariants is
\begin{subequations}
\begin{eqnarray}
\hat{s}&=& (k_1+k_2)^2,
\\
\hat{s}_1&=& (k_1+k_2-P)^2,
\\
Y &=& (k_1\cdot P)(k_2\cdot P).
\end{eqnarray}
\end{subequations}
Making use of Lorentz invariance, we can evaluate the $t\bar t$ 2-body phase-space integral 
in the rest frame of $t\bar t$, with the Higgs momentum in the $z$ direction. 
We set $M_H=0$ in the calculation of $\td\hat{\sigma}$,
which greatly simplifies the phase-space integration. 
The individual terms in the differential cross section in Eq.~\eqref{eq:Httbar0} are
\begin{eqnarray}
\label{eq:Httbar}
\frac{\td^2\hat{\sigma}^{\text{ij} }}{\td P_T^2 \td \hat y} =
\frac{\as^2\, y_t^2}{36\,\pi\,\hat{s}^{\,3}\hat{s}_1(\hat{s}-\hat{s}_1)^4}
\bigg\{
C^{\text{ij}}(\hat{s},\hat{s}_1,Y) \log \frac{(\sqrt{\hat{s}_1}+\sqrt{\hat{s}_1-4M_t^2}\,)^2}{4M_t^2} 
\nonumber\\
+D^{\text{ij}}(\hat{s},\hat{s}_1,Y) \sqrt{\hat{s}_1}\sqrt{\hat{s}_1-4M_t^2}
\bigg\}.
\end{eqnarray}
The functions $C^{\text{ij} }$ and $D^{\text{ij} }$
are polynomials in the Lorentz invariants $\hat{s}$, $\hat{s}_1$, and $Y$ and in $M_t^2$.
In the contribution to the cross section from diagram a, the functions are
\begin{subequations}
\begin{eqnarray}
C^{\text{aa}}&=&
\hat{s}\hat{s}_1(\hat{s}-\hat{s}_1)^2\left[ (\hat{s}-\hat{s}_1)^2-8\, Y\right]
-8\,\hat{s}\hat{s}_1 M_t^2 \left[\hat{s}_1(\hat{s}-\hat{s}_1)^2-8\,Y(\hat{s}+2\hat{s}_1)\right],
\\
D^{\text{aa}}&=&
-\hat{s} (\hat{s}-\hat{s}_1)^2 (\hat{s}^2-6\,\hat{s} \hat{s}_1+\hat{s}_1^2)
+8\,Y(\hat{s}^3-5\hat{s}^2\hat{s}_1-\hat{s} \hat{s}_1^2+\hat{s}_1^3)
\nonumber\\
&&
+16\,M_t^2\left[\hat{s} \hat{s}_1(\hat{s}-\hat{s}_1)^2-2\,Y(\hat{s}^2+6\hat{s} \hat{s}_1+\hat{s}_1^2)\right].
\end{eqnarray}
\end{subequations}
In the contribution to the cross section from
the interference between  diagrams a  and b, the functions are
\begin{subequations}
\begin{eqnarray}
C^{\text{ab}}&=&
8\,\hat{s} \hat{s}_1 M_t^2\left[\hat{s}(\hat{s}-\hat{s}_1)^2-8\, Y \hat{s}_1\right]
+16\hat{s} \hat{s}_1 M_t^4 \left[(\hat{s}-\hat{s}_1)^2-8\,Y\right],
\nonumber\\
D^{\text{ab}}&=&
\hat{s}(\hat{s}-\hat{s}_1)^4-8\, Y(\hat{s}-\hat{s}_1)^2 (\hat{s}+\hat{s}_1)
\nonumber\\
&&
-8\,M_t^2\left[\hat{s} \hat{s}_1(\hat{s}-\hat{s}_1)^2-4\,Y(\hat{s}^2+4\hat{s} \hat{s}_1+\hat{s}_1^2)\right].
\end{eqnarray}
\end{subequations}
Substituting Eqs.~\eqref{eq:Httbar} into Eq.~\eqref{eq:Httbar0} 
and subtracting the fragmentation term in Eq.~\eqref{eq:FragCont},
we obtain the infrared-safe differential cross section for $q\bar q\to H+t\bar t$.

The only mass singularity in Eq.~\eqref{eq:Httbar} is  in the $C^{\textrm{aa}}$  term. 
The singular part  is
\begin{equation}
\frac{\td^2\hat{\sigma}^{\text{aa} }}{\td P_T^2 \td \hat y}\bigg|_{\text{singular}}
=
\frac{\as^2\, y_t^2}{36\,\pi\,\hat{s}^2(\hat{s}-\hat{s}_1)^2}
\left[(\hat{s}-\hat{s}_1)^2-8\,Y\right]
\log \frac{\Big(\sqrt{\hat{s}_1}+\sqrt{\hat{s}_1-4M_t^2}\Big)^2}{4M_t^2}.
\end{equation}
The logarithm of $1/M_t^2$ becomes a mass singularity in the limit $M_t \to 0$.
By using $Y/(\hat{s}-\hat{s}_1)^2 = (1-\tanh^2 \hat y)/16$, we see that the
coefficient of the mass singularity  is the same as in 
the fragmentation term in Eq.~\eqref{eq:FragCont}. 
The differential cross section 
${\td^2\tilde{\sigma}_{q\bar{q}\rightarrow H+t\bar t }}/{\td P_T^2 \td \hat y}$
defined by the subtraction in Eq.~\eqref{eq:SDC2} is therefore infrared safe. 

\subsection{Results}

We now compare the  LP factorization formula at LO
with the complete LO result for the
contribution to inclusive Higgs production at a 100~TeV $p p$ collider
 from the specific partonic process $q\bar{q}\to H t\bar t$. 
Our goal is  not to give quantitative predictions for Higgs production in $pp$ collisions,
which is dominated by $gg\to H t\bar t$ and $gq \to H t\bar t$,
but rather to estimate the minimum Higgs transverse momentum $P_T$ 
above which the LP factorization formula is reliable. 
Neither do we consider the top-quark initiated process, 
which might have non-ignorable contribution \cite{Han:2014nja}.  
A thorough treatment of the Higgs $P_T$ distribution in $p p$ collisions 
using the LP factorization formula at LO will be presented in a future work.

We consider the double differential cross section in the transverse momentum $P_T$ 
and rapidity $y$ of the Higgs.
The LP factorization formula at LO for the contribution from $q\bar{q}\to H t\bar t$ 
is given in Eq.~\eqref{eq:LPfacformula_LO}.
The rapidity $y$ of the Higgs in the $pp$ center-of-momentum frame
is related to its rapidity $\hat y$ in the $q\bar q$ center-of-momentum frame by 
\begin{equation}
y=\hat y+\frac{1}{2}\log(x_a/x_b),
\end{equation}
where $x_a$ and $x_b$ are the longitudinal momentum fractions of the colliding partons
from the protons $A$ and $B$.  We use the fragmentation function for $t \to H$ 
in the $\overline{\text{MS}}$ factorization scheme  in Eq.~\eqref{eq:fac_formula_LO_PT}, 
and we consider both the ZMTQ and hybrid factorization prescriptions.
We do not consider the resummation of large logarithms,  
so the $t \to H$ fragmentation function depends only linearly on the 
logarithm of the fragmentation scale $\mu$.

\begin{figure}
\begin{center}
\includegraphics[width=0.9\textwidth]{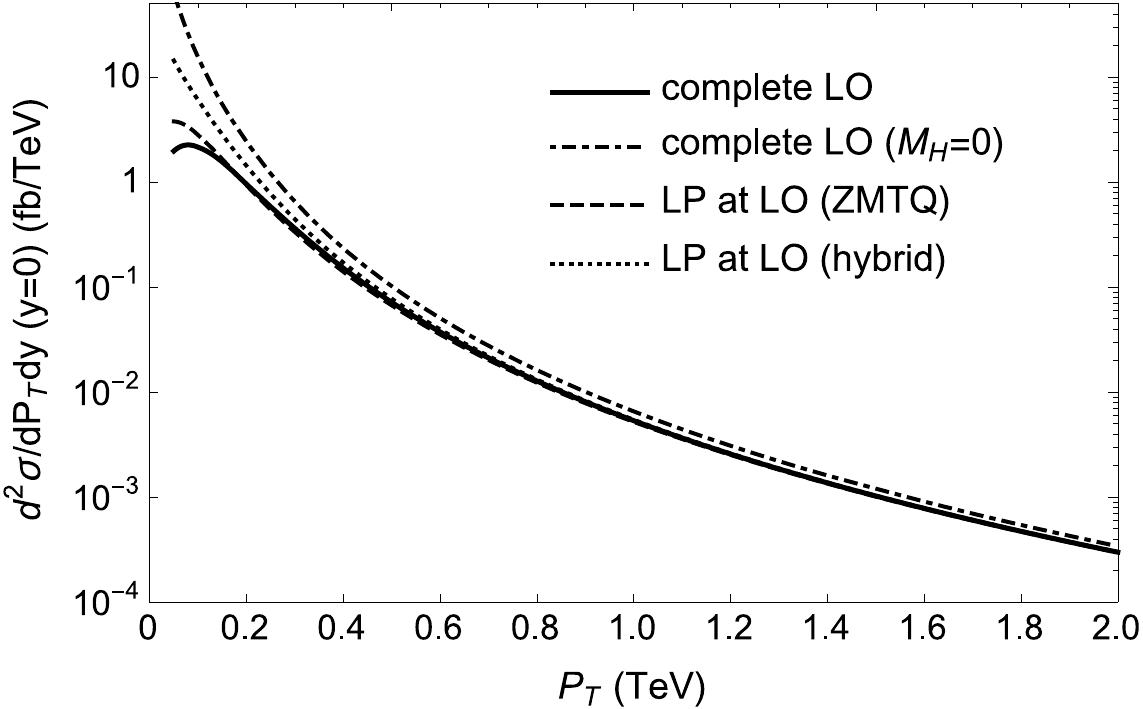}
\caption{ 
Differential cross section for inclusive Higgs production at central rapidity
from the parton process $q\bar{q}\to t\bar{t}H$ at a 100~TeV $p p$ collider
as a function of the Higgs transverse momentum $P_T$. 
The complete LO result (solid curve) is compared to three approximations:
the complete LO with $M_H=0$ (dot-dashed curve),
the LP factorization at LO using the ZMTQ prescription (dashed curve), 
and the LP factorization at LO using the hybrid prescription (dotted curve).
The LP factorization results are calculated using the fragmentation function for $t \to H$
in the $\overline{\text{MS}}$ scheme.
\label{fig:PT}}
\end{center}
\end{figure}

In Fig.~\ref{fig:PT}, we compare the complete LO result for the transverse momentum distribution 
of the Higgs boson at rapidity $y=0$,
which is given in Appendix~\ref{app:LO}, with three approximations:  
the complete LO cross  section with $M_H=0$, 
the LP factorization formula at LO using
the ZMTQ prescription,
and the LP factorization formula at LO using the hybrid prescription.
The $\overline{\text{MS}}$ scheme is used for the $t\to H$ fragmentation function.
We consider $pp$ collisions with center-of-mass energy $\sqrt{s}=100$~TeV. 
We use CTEQ6.6M parton distributions \cite{Nadolsky:2008zw}
and $\as(M_Z)=0.13$, both with $n_f=5$. 
We choose the factorization scale and the fragmentation scale to both be $\mu=\sqrt{P_T^2+M_H^2}$.
The four curves in Fig.~\ref{fig:PT} are
\begin{itemize}
\item 
{\it complete LO}:
complete LO cross section
with $M_t=173$~GeV and $M_H=125$~GeV. 
\item 
{\it complete LO ($M_H=0$)}:
complete LO cross section 
with $M_t=173$~GeV and $M_H=0$. 
\item 
{\it LP at LO (ZMTQ)}: 
LP factorization formula at LO using the ZMTQ prescription
in which $M_t=0$ in the infrared-safe cross section 
for $q\bar{q}\rightarrow t +\bar  t$ and $M_H=M_t=0$
 in the infrared-safe cross section for $q\bar{q}\rightarrow H + t\bar  t$.
\item 
{\it LP at LO (hybrid)}:
LP factorization formula at LO using the hybrid prescription
in which $M_t=0$ in the infrared-safe cross section 
for $q\bar{q}\rightarrow t +\bar  t$ and $M_H=0$ and
$M_t=173$~GeV in the infrared-safe cross section for $q\bar{q}\rightarrow H + t\bar  t$.
\end{itemize}
In both the ZMTQ and hybrid prescriptions, the masses
in the $t \to H$ fragmentation function  
are $M_H=125$~GeV and $M_t=173$~GeV.

Fig.~\ref{fig:PT} shows that 
the LP factorization formula with the ZMTQ and hybrid prescriptions
 both give increasingly good approximations  to the complete LO result at large $P_T$,
with the errors decreasing to below $5\%$ for $P_T > 600$ GeV.
 In contrast, the fractional error for the complete LO result 
with a massless Higgs does not go to zero at large $P_T$.
The reason is that when the complete LO result in Appendix~\ref{app:LO} 
is expanded in powers of the softer scale $Q_S$ ($M_H$ or $M_t$) 
divided by the harder scale $Q_H$ ($P_T$ or $\sqrt{\hat s}$), 
the leading-power term includes 
\begin{equation}\label{eq:LPterm}
\begin{split}
&\frac{\as^2 \,y_t^2}{36\,\pi\hat{s}^{\,2}}(1+\tanh^2 \hat{y})
\bigg\{\log\frac{\hat{s}}{4M_t^2}
-\log\frac{\cosh^2 \hat{y}\, P_T^2+\zeta_t\, \sqrt{\hat s}\,(\sqrt{\hat s}-2\cosh\hat{y}\,P_T)}
              {4\cosh^2  \hat{y} \, P_T^2\,(\sqrt{\hat s}-2\cosh \hat{y}\, P_T)/\sqrt{\hat s}}
\\&\hspace{4cm}
+
\frac{(1-\zeta_t)\sqrt{\hat s}\,(\sqrt{\hat{s}}-2  \cosh \hat y\, P_T)}
{\cosh^2 \hat y\, P_T^2+ \zeta_t\,\sqrt{\hat s}\,(\sqrt{\hat s}-2 \cosh \hat y \,P_T)}
\bigg\},
\end{split}
\end{equation}
where $\zeta_t=M_H^2/(4M_t^2)$.
Setting $M_H=0$ in this expression leads to an error of order $M_H^2/M_t^2$
that does not decrease with increasing $P_T$. 
Therefore the complete LO result with $M_H = 0$
does not converge to the complete LO result at large $P_T$.
The expression in curly brackets in Eq.~\eqref{eq:LPterm} differs from that in
the fragmentation contribution in Eq.~\eqref{eq:FragCont} only by
a logarithm whose argument depends on $\hat s$, $P_T$, and $\hat y$ 
but does not depend on $M_H$ or $M_t$. 
Consequently, the terms suppressed by only  $M_H^2/M_t^2$ cancel in the infrared-safe cross section 
for $q\bar{q}\to H+t\bar t$, which is defined in Eq.~\eqref{eq:tildesigma-Htt_LO}
by subtracting the fragmentation contribution from the complete LO result.
Therefore, setting $M_H=0$ in this infrared-safe cross section only gives errors of order $M_H^2/P_T^2$.

\begin{figure}
\begin{center}
\includegraphics[width=0.9\textwidth]{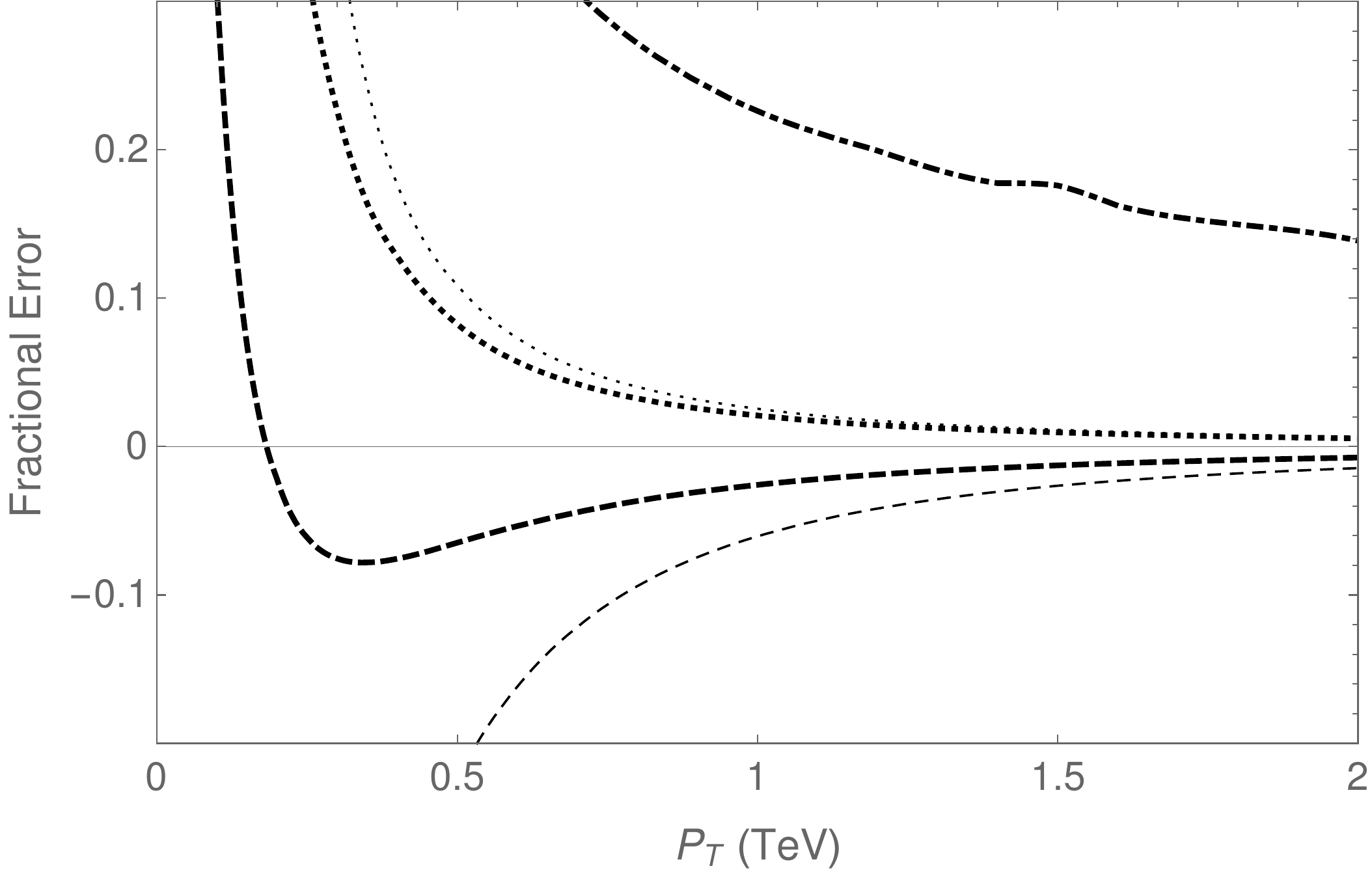}
\caption{
 Fractional error in the  
differential cross section for inclusive Higgs production with five approximations: 
complete LO with $M_H=0$ (dot-dashed curve), 
LP factorization at LO using the ZMTQ prescription (dashed  curves), 
and LP factorization at LO using the hybrid prescription (dotted curves).
The LP factorization results are calculated using the fragmentation function for $t \to H$
in the $\overline{\text{MS}}$ scheme (thicker curves) and in the IMC scheme (thinner curves).
\label{fig:Err}}
\end{center}
\end{figure}

In Fig.~\ref{fig:Err}, we show the fractional errors for five approximations to the complete LO
differential cross section as functions of $P_T$.
The fractional error is the difference between the approximate result 
and the complete LO result divided by the complete LO result.
The fractional error does not go to zero at large $P_T$ for 
the complete LO result with $M_H=0$, but it does go to zero for the LP factorization results at LO 
with either the ZMTQ or hybrid prescriptions  and with either the $\overline{\text{MS}}$ or IMC schemes.
For both the $\overline{\text{MS}}$ and IMC schemes, 
the LP factorization cross section approaches the complete LO cross section from 
below with the ZMTQ prescription and from above with the hybrid prescription.
The convergence to zero is faster for the $\overline{\text{MS}}$ scheme than for the IMC scheme. 
Thus the more physical phase space constraints 
in the IMC fragmentation function does not lead to faster convergence.
In the $\overline{\text{MS}}$ scheme,
the fractional errors at large $P_T$  for both factorization prescriptions are numerically
consistent with the simple estimate $M_H^2/P_T^2$.
The predicted fractional  errors
are order $M_H^2/P_T^2$ with the hybrid prescription,
but they are order $M_t^2/P_T^2$ with the ZMTQ prescription.
For the ZMTQ prescription, the fractional error is considerably smaller than the 
simple estimate $M_t^2/P_T^2$.  It is interesting that the ZMTQ prescription, 
in which the top-quark mass is set to zero in the infrared-safe cross section
for $q\bar{q}\rightarrow H + t\bar  t$, gives essentially the same numerical convergence 
to the complete LO result as the hybrid prescription,
despite having a parametrically larger error.


\section{Summary and  Outlook}
\label{sec:conc}

We have presented the leading-power (LP) factorization formula for Higgs production 
with transverse momentum $P_T$  much larger than the mass $M_H$ of the Higgs.
In hard-scattering cross sections for Higgs production,
all terms with the leading power of  $1/P_T^2$ are expressed as convolutions of 
infrared-safe cross sections for producing a fragmenting parton
and fragmentation functions that give the distribution of  longitudinal momentum of the 
Higgs in the jet produced by the fragmenting parton.
The LP factorization formula separates the scales $P_T$ and $M_H$,
with all the dependence on $P_T$ being in the infrared-safe cross sections
and all the dependence on $M_H$ being in the fragmentation functions.
The errors in the LP factorization formula are order $M_H^2/P_T^2$.

The fragmentation functions for Higgs production can be calculated diagrammatically 
as expansions in powers of coupling constants.
In contrast to the fragmentation functions for hadron production in QCD,
the fragmentation functions for Higgs production are completely perturbative.
The fragmentation functions for $W$ and $Z$ into Higgs
and for top quark into Higgs were calculated at LO in the Standard Model coupling constants
using two factorization schemes:  the $\overline{\textrm{MS}}$ scheme and the 
invariant-mass cutoff (IMC) scheme.
The $\overline{\textrm{MS}}$ scheme can be defined to all orders in the coupling constants, 
while  the IMC scheme is defined only at LO.
In the IMC scheme, the factorization scale $\mu$ has a physical interpretation 
as the maximum invariant mass of a jet that includes the Higgs.
If the factorization scale  is much larger than $M_H$,
the IMC fragmentation function with factorization scale $\mu$
can be approximated by the corresponding $\overline{\textrm{MS}}$ fragmentation function
with a factorization scale $\mu/\sqrt{z(1-z)}$ multiplied by a constant
that is different for $W,Z$ into Higgs and for top quark into Higgs.
The fragmentation functions satisfy evolutions equations that can be used to sum
leading logarithms of $\mu/M_H$ to all orders in the coupling constants.
The splitting functions in the evolution equations can be calculated perturbatively.
Leading logarithms of $P_T/M_H$ in the cross section can be summed to all orders
by choosing the fragmentation scale $\mu$ to be of order $P_T$.
The summation of these logarithms can improve upon the accuracy of fixed-order calculations.

The infrared-safe cross sections can be calculated diagrammatically order-by-order in the 
coupling constants by subtracting mass singularities from the hard-scattering cross sections.
This procedure was carried out explicitly for the LP factorization formula for 
$q \bar q \to H t \bar t$ at LO.  We considered three factorization prescriptions
for the infrared-safe cross sections: the massive-top-quark (MTQ) prescription,
the zero-mass-top-quark (ZMTQ) prescription, and a hybrid prescription.
The  fractional errors are order $M_H^2/P_T^2$ for the MTQ  and hybrid prescriptions
and order $M_t^2/P_T^2$ for the ZMTQ prescription.
The ZMTQ and hybrid prescriptions have the advantage 
of a simple relation between the momenta of the fragmenting top quark 
and the Higgs:  $p = \tilde{P}/z$, where $\tilde{P}$ is light-like.
In the ZMTQ prescription, the only scale in the infrared-safe cross sections
is the transverse momentum $p_T$ of the fragmenting parton,
which greatly simplifies the calculation of higher-order corrections.

The LP factorization formula was illustrated by calculating the contribution
to Higgs production at a 100~TeV $pp$ collider from the subprocess $q \bar q \to H t \bar t$.
The LP factorization formula at LO was compared to the complete LO result
for the $P_T$ distribution at central rapidity.
The fractional error in the LP factorization formula decreases at large $P_T$ like $1/P_T^2$.
For the ZMTQ and hybrid prescriptions using the $t \to H$ fragmentation function 
in the $\overline{\textrm{MS}}$  scheme,
the fractional error decreases to less than 5\% for $P_T < 600$~GeV.
The fractional errors are larger for the $t \to H$ fragmentation function in the IMC  scheme.
Thus the physical phase-space constraints in the IMC fragmentation function
do not lead to faster convergence with $P_T$.
With the $\overline{\textrm{MS}}$ fragmentation function,
the fractional error for the ZMTQ prescription is numerically approximately equal to that for the hybrid prescription.
This is surprising given that the theoretical error for the ZMTQ prescription
is parametrically larger: order $M_t^2/P_T^2$ compared to $M_H^2/P_T^2$ for the hybrid prescription.
The smaller theoretical error in the hybrid prescription is obtained by taking into account
the top quark mass in some of the infrared-safe cross sections,
which makes the calculations much more difficult.
Our results for the specific subprocess $q \bar q \to H t \bar t$ suggest that the smaller theoretical error
may not be worth the additional calculational effort.

We have calculated the infrared-safe cross sections in the LP factorization formula at LO 
only for the hard-scattering process $q \bar q \to H t \bar t$.
It is straightforward to calculate  the infrared-safe cross sections 
for the other hard-scattering processes at order $\alpha_s y_t^2$, such as $g g \to H t \bar t$.
The only fragmentation function that is needed in order to subtract the mass singularities
is the $t \to H$ fragmentation function at LO, which we have calculated in this paper.
A phenomenologically relevant application of the LP factorization formula at LO
to inclusive Higgs production in association with a $t \bar t$ pair requires 
that these other subprocesses be included.
It would be interesting to compare the errors in such a calculation 
for the ZMTQ and hybrid factorization prescriptions.
If the errors are numerically comparable for ZMTQ in spite of being parametrically larger,
it would further strengthen the case for using the ZMTQ prescription at higher orders.
Significantly more calculational effort would be required to apply the LP factorization formula 
to inclusive Higgs production in association with a $t \bar t$ pair at NLO.
The advantage over the complete NLO calculation is that the  theoretical errors could be further decreased 
by using the evolution equations for the fragmentation functions to sum the leading logarithms of $P_T/M_H$.

The LP factorization formula can also be applied to inclusive Higgs production 
without $t \bar t$.
The most important couplings of Higgs $P_T$ below the $t \bar t$ threshold
are its couplings to gluons in the effective field theory (HEFT)
obtained by integrating out top quark loops.
In the LP factorization formula at LO, the only fragmentation function is the $H \to H$
fragmentation, which is given by the delta function in Eq.~\eqref{eq:DHH}.
The infrared-safe cross sections are just the hard-scattering cross sections with $M_H$ set to zero.
Thus the LP factorization formula at LO is just the cross section for producing a massless Higgs.
The LP factorization formula at NLO involves the LO fragmentation function for $g \to H$,
which comes from the tree-level process $g^* \to Hg$ 
through the HEFT vertex that couples the Higgs to two gluons.
Some of the infrared-safe cross sections at this order must be obtained by the subtraction 
of mass singularities.  
The LP factorization formula at N$^2$LO in HEFT would be much more difficult to calculate.
The complete N$^2$LO cross section  in HEFT as a function of $P_T$
has already been calculated \cite{Boughezal:2015dra,Boughezal:2015aha}.  
The fractional error  of the LP factorization formula at N$^2$LO
relative to the complete  N$^2$LO cross section is order $M_H^2/P_T^2$.
Given that the applicability of the LP factorization formula in HEFT is limited to $P_T$
below the top-quark-pair threshold, logarithms of $P_T/M_H$ cannot be very large.
Thus the LP factorization formula at N$^2$LO in HEFT cannot improve 
significantly upon the accuracy of the complete N$^2$LO calculation in HEFT by summing logarithms.  
However it does have the advantage of greater simplicity 
provided by the separation of the scales $P_T$ and $M_H$.  
Thus it may provide physical insights into the results of the complete N$^2$LO calculation.

Factorization theorems for inclusive quarkonium production at large $P_T$
have been extended to the next-to-leading power (NLP) in $1/P_T^2$
\cite{Kang:2011mg,Fleming:2012wy,Kang:2014tta,Kang:2014pya}.
The NLP factorization formula involves new production mechanisms called 
double-parton fragmentation, in which the quarkonium is produced in a jet that results 
from the hadronization of two collinear partons produced in a hard collision.
The NLP factorization formula for inclusive quarkonium production in QCD
can be adapted straightforwardly to inclusive Higgs production in the Standard Model.
For $P_T$ above the top-quark-pair threshold, double-parton fragmentation 
first enters at LO, which is order $\alpha_s^2 y_t^2$, through the fragmentation function 
for $t \bar t \to H$.
The fractional error in the NLP factorization formula is order $M_H^4/P_T^4$.
The simple estimate $M_H^4/P_T^4$ decreases to about 6\% at $P_T = 250$~GeV.
Thus the NLP factorization formula could be useful even at the Large Hadron Collider.

The LP factorization formula could be useful for quantifying the effects of physics 
beyond the Standard Model on Higgs production at large $P_T$
\cite{Azatov:2013xha,Grojean:2013nya,Schlaffer:2014osa,Dawson:2015gka}.
The new physics would modify both the fragmentation functions and the infrared-safe cross sections.
This could be important if the fraction of the cross section for Higgs production
from new physics is much larger at large transverse momentum.

     
\acknowledgments
This work was supported in part by the Department of Energy under grant DE-SC0011726.
HZ would like to thank Jian-Wei Qiu, Richard Furnstahl, Fredrick Olness, and Yan-Qing Ma 
for beneficial discussions.


\appendix 


\section{Feynman rules for fragmentation functions}
\label{app:FRules}

Fragmentation functions can be calculated using  Feynman rules
derived by Collins and Soper in 1981~\cite{Collins:1981uw}.
The fragmentation function is expressed as the sum of all possible cut diagrams
of a particular form.  The diagrams have an eikonal line that
extends from the vertex of a local operator on the left side of the cut
to the vertex of a local operator on the right side.
The virtual parton lines attached to the operator vertices are connected to the fragmented particle 
through ordinary field theory interactions, with 
possibly additional parton lines attached to the eikonal line.
The cut passes through the eikonal line, the line for the fragmented particle,
and possibly additional lines that correspond to additional final-state particles.
Example of cut diagrams for producing a Higgs boson
are shown in Figures~\ref{fig:V->H_FD} and \ref{fig:t->H_FD}.
In these figures,
there is a Higgs line that ends on each side of the cut. 
It is not drawn as passing through the cut to emphasize that its momentum is not integrated over,
unlike the other cut parton lines.

The Feynman rules for the cut diagrams are relatively simple \cite{Collins:1981uw}.
The 4-momentum $K$ of the fragmenting parton
enters the diagram through the operator vertex on the left side of the eikonal line
and it exits through the operator on the right side.
Some of that momentum flows through the virtual partons attached to the operator
vertex and the remainder flows through the eikonal line.
The fragmented particle, which in this case is a Higgs, has a specified 4-momentum $P$.
The longitudinal momentum fraction $z$ of the Higgs  is 
$ z =P\cdot n/K \cdot n$, where $n$ is a light-like 4-vector.
The fragmentation function depends on $K$
only through $z$.

The local operator vertices in the Feynman diagram are connected by an eikonal line.
The propagator for the eikonal line is $i/(q\cdot n+ i \epsilon)$,
where $q$ is the momentum flowing through the eikonal line.
The Feynman rule for the cut eikonal line is $2 \pi \delta(q \cdot n)$.
In QCD, the eikonal factor is an operator that corresponds to a path-ordered exponential 
of gluon fields in an appropriate color representation.
Thus there are diagrams with eikonal vertices at which gluon lines attach to the eikonal line.
A gluon attached to the eikonal line has a Lorentz index $\beta$ and color-octet index $c$. 
In a gluon fragmentation function, 
the propagator for an eikonal line carrying momentum $q$ on the left side of the cut
is $i \delta^{de}/(q \cdot n+ i \epsilon)$ 
and the Feynman rule for the eikonal vertex on the left side of the cut is $g_s f^{cde}n^\beta$, 
where $d$ and $e$ are the color-octet indices to the left and right of the propagator or vertex.
In a quark fragmentation function, 
the propagator for an eikonal line carrying momentum $q$ on the left side of the cut
is $i \delta_{ji}/(q.n+ i \epsilon)$ 
and the Feynman rule for the eikonal vertex on the left side of the cut is $i g_s T^c_{ji} n^\beta$, 
where $i$ and $j$ are the color-triplet indices  to the left and right of the propagator or vertex.

For a gluon fragmentation function, the local operators in the definition of the fragmentation function
are $n^\sigma G^b_{\sigma \mu}$, where $G^b_{\sigma \mu}$ is the gluon field strength.
The Feynman diagrams can be rearranged in such a way that the 
operator creates only a single virtual-gluon line~\cite{Collins:1981uw}.
The operator vertex at the left end of the eikonal line is labelled by a Lorentz index
$\mu$ and an color-octet index $c$.
If the single virtual-gluon line attached to that operator 
has outgoing momentum $q$, Lorentz index $\alpha$, and color-octet index $a$,
the Feynman rule for the operator vertex is
\begin{equation}
\label{eq:gvertex}
-i(K\cdot n \,g^{\mu \alpha} - q^\mu n^\alpha) \delta^{ab}.
\end{equation} 
The operator vertex at the right end of the eikonal line is labelled by a Lorentz index
$\nu$ and a color-octet index $c$.  
The gluon fragmentation function is the sum of all cut diagrams 
contracted with~\cite{Collins:1981uw}
\begin{equation}
\label{eq:goverall}
\frac{z^{D-3}}{(N_c^2-1)(D-2)2\pi K\cdot n} (-g_{\mu\nu})\delta^{de},
\end{equation} 
where $N_c$ is the number of  quark colors, $D$ is the number of spacetime dimensions,
and $d$ and $e$ are the color-octet indices of the cut eikonal propagator.
The factors in the denominator include the $N_c^2-1$ color states 
and  the $D-2$ physical spin states of a gluon.
The factor of $z^{D-3}$ arises from an integral over a transverse momentum.

For a quark fragmentation function in QCD, the local operators  
in the definition of the fragmentation function are the quark field operator $\bar \psi$
and its hermitian conjugate.
The operator vertex at the left end of the eikonal line can be labelled by a Dirac index
and a color-triplet index, but it is more convenient to leave those indices implicit.
The quark fragmentation function is the trace in Dirac indices and in color-triplet indices
of the sum of all cut diagrams 
multiplied by~\cite{Collins:1981uw}
\begin{equation}
\label{eq:qoverall}
\frac{z^{D-3}}{8 N_c \pi K.n}  n \cdot \gamma.
\end{equation} 
The factors in the denominator include the $N_c$ color states 
and the 2 physical spin states of a quark.
The suppressed Dirac indices of the matrix $n \cdot \gamma$
are contracted with a Dirac index of the propagator of the virtual quark 
created by the operator vertex on the left and a Dirac index of the propagator of the virtual quark 
absorbed by the operator vertex on the right.
There is also an implicit unit color matrix in Eq.~\eqref{eq:qoverall}
whose color-triplet indices are contracted with those of the cut eikonal line.

For a weak vector boson fragmentation function, 
the local operators in the definition of the fragmentation function
are $n^\sigma F_{\sigma \mu}$, where $F_{\sigma \mu}$ is the field strength
for the vector boson.
The Feynman rule for the operator vertex is the same as in Eq.~\eqref{eq:gvertex},
except that the color factor $\delta^{ac}$ is omitted.
The fragmentation function is the sum of all cut diagrams 
contracted with a factor that can be obtained from Eq.~\eqref{eq:goverall}
by omitting the factors $\delta^{de}/(N_c^2-1)$.


\section{Complete LO result for $\bm{q\bar{q}\to H t \bar t}$}
\label{app:LO}

In this appendix, we present the complete LO cross section for $q\bar{q}\to H t \bar t$.
The two Feynman diagrams are shown in Figure~\ref{fig:FG},
and they are labelled a and b.
We denote the momenta of the $q$, $\bar q$ and $H$
by $k_1$, $k_2$ and $P$, respectively.
We express the cross section in terms of the following Lorentz invariants
\begin{eqnarray}
\s&=& (k_1+k_2)^2,
\\
\sa&=& (k_1+k_2-P)^2,
\\
Y&=&(k_1\cdot P)(k_2\cdot P).
\end{eqnarray}
We also use the notation $\lambda(a,b,c)=a^2+b^2+c^2-2ab-2ac-2bc$.

The complete LO result for process for $q\bar{q}\to H t \bar t$ can be expressed 
in the same manner as in Eq.~\eqref{eq:Httbar0}:
\begin{equation}
\frac{\td^2\hat{\sigma}_{q\bar{q}\rightarrow H+t\bar t }}{\td P_T^2 \td \hat y}
=
2\left(\frac{\td^2\hat{\sigma}^{\text{aa} }}{\td P_T^2 \td \hat y}
+\frac{\td^2\hat{\sigma}^{\text{ab} }}{\td P_T^2 \td \hat y}\right).
\end{equation}
The contribution to the cross section from diagram a is
\begin{eqnarray}
\frac{\td^2\hat{\sigma}^{\text{aa} }}{\td P_T ^2\td \hat y}
&=&
\frac{\as^2\, y_t^2}{36\,\pi\,\hat{s}^{\,3}\hat{s}_1\,
\lambda^{5/2}(\hat{s},\hat{s}_1,M_H^2)\, 
}
\nonumber\\
&&\times
\bigg\{
C^{\text{aa}}(\hat{s},\hat{s}_1,Y)\, \hat{s}\,\hat{s}_1
\,\log\frac{\big[ (\hat{s}-\hat{s}_1+M_H^2) \hat{s}_1^{1/2} 
                        + (\hat{s}_1-4M_t^2)^{1/2}\, \lambda^{1/2}(\hat{s},\hat{s}_1,M_H^2) \big]^2}
               {4\big[M_H^2 \hat{s} \hat{s_1}+M_t^2\lambda(\hat{s},\hat{s}_1,M_H^2)\big]}
\nonumber\\
&&\hspace{1cm}
-D^{\text{aa}}(\hat{s},\hat{s}_1,Y) \,
\frac{\hat{s}_1^{1/2} (\hat{s}_1-4M_t^2)^{1/2}\, \lambda^{1/2}(\s,\sa,M_H^2)}
       {M_H^2\hat{s}\hat{s}_1 +M_t^2\lambda(\hat{s},\hat{s_1},M_H^2)}
\bigg\}.
\end{eqnarray}
The functions $C^{\text{aa}}$ and $D^{\text{aa}}$
are polynomials in the Lorentz invariants $\hat{s}$, $\hat{s}_1$, and $Y$ 
and in the masses $M_t^2$ and $M_H^2$.
They can be expanded in powers of $M_H^2$:
\begin{subequations}
\begin{eqnarray}
C^{\text{aa}}(\hat{s},\hat{s}_1,Y)&=& 
\sum_{n} C^{aa}_{n}(\hat{s},\hat{s}_1,Y)  (M_H^2)^n,
\\
D^{\text{aa}}(\hat{s},\hat{s}_1,Y)&=&
\sum_{n} D^{aa}_{n}(\hat{s},\hat{s}_1,Y)  (M_H^2)^n.
\end{eqnarray}
\label{eq:expandMH}
\end{subequations}
The expansion coefficients  for $C^{\text{aa}}(\hat{s},\hat{s}_1,Y)$ are
\begin{subequations}
\begin{eqnarray}
C^{\text{aa}}_{0}&=&
(\hat{s}-\hat{s}_1)^3\left[(\s-\sa)^2-8M_t^2 \sa\right]-8Y(\s-\sa)\left[(\s-\sa)^2-8M_t^2(\s+2\sa)\right],
\\
C^{\text{aa}}_{1}&=&
-(\s-\sa)\left[(\s-\sa)^2(\s+3\sa)+8M_t^2(2\s^2+3\s\sa-3\sa^2)\right]
\\
&&
+8Y\left[(\s-\sa)(\s-3\sa)-8M_t^2(2\s-\sa)\right],
\\
C^{\text{aa}}_{2}&=&
4\left[\sa(2\s^2-\s\sa-\sa^2)+M_t^2(8\s^2-2\s\sa+6\sa^2)\right]
+8Y(\s-3\sa+8M_t^2),
\\
C^{\text{aa}}_{3}&=& 4(2\s+\sa)(\sa-2M_t^2)-8Y,
\\
C^{\text{aa}}_{4}&=& -(\s+3\sa),
\\
C^{\text{aa}}_{5}&=& 1,
\end{eqnarray}
\end{subequations}
The expansion coefficients for $D^{\text{aa}}(\hat{s},\hat{s}_1,Y)$ are
\begin{subequations}
\begin{eqnarray}
D^{\text{aa}}_{0} &=&
M_t^2 \s (\s-\sa)^4 \left[\s^2-6\s\sa+\sa^2-16M_t^2\sa\right]
\nonumber\\
&&
-8Y M_t^2(\s-\sa)^2\left[ \s^3-5\s^2\sa-\s\sa^2 +\sa^3-4M_t^2(\s^2+6\s\sa+\sa^2) \right],
\\
D^{\text{aa}}_{1}&=&
2\s(\s-\sa)^2 \left[\s\sa(\s-\sa)^2-2M_t^2\s(\s^2-4\s\sa+\sa^2)-4M_t^4(\s^2-2\s\sa-7\sa^2)\right]
\nonumber\\
&&
-8Y \left[\s\sa(\s-\sa)^2(2\s+\sa)-4M_t^2(\s^4-3\s^3\sa+10\s^2\sa^2-\s\sa^3+\sa^4) \right.
\nonumber\\
&& \hspace{1cm}
\left. +16M_t^4 (\s+\sa)^3\right],
\\
D^{\text{aa}}_{2} &=&
-\s \left[2\s\sa(\s-\sa)^2(2\s+\sa)-M_t^2(7\s^4-28\s^3\sa-54\s^2\sa^2+20\s\sa^3-9\sa^4) \right.
\nonumber\\
&& \hspace{1cm}
\left. -32M_t^4(\s^3+\s\sa^2-2\sa^3)\right]
+16Y \left[\s\sa(2\s^2-3\s\sa+\sa^2) \right.
\nonumber\\
&& \hspace{1cm}
\left. -M_t^2(3\s^3-\s^2\sa+\s\sa^2+3\sa^3)  +12M_t^4 (\s+\sa)^2\right],
\\
D^{\text{aa}}_{3} &=&
4\s \left[ \s^2\sa(\s+3\sa)-M_t^2(2\s^3-6\s^2\sa-2\s\sa^2-4\sa^3)-4M_t^4(3\s-\sa)(\s+\sa)\right]
\nonumber\\
&&
-8Y \left[\s\sa(2\s+\sa)-4M_t^2(\s^2+\s\sa+\sa^2)+16M_t^4 (\s+\sa)\right],
\\
D^{\text{aa}}_{4} &=&
-\s \left[2 \s \sa(2\s+\sa)-M_t^2(7\s^2-10\s\sa-9\sa^2)-16M_t^4(2\s+\sa)\right]
\nonumber\\
&&
-8Y M_t^2 (\s+\sa-4M_t^2),
\\
D^{\text{aa}}_{5} &=& 2\s(\s\sa-2M_t^2\s-4M_t^4),
\\
D^{\text{aa}}_{6} &=& M_t^2\s,
\end{eqnarray}
\end{subequations}

The contribution to the cross section from
the interference between  diagrams a  and b is
\begin{eqnarray}
\frac{\td^2\hat{\sigma}^{\text{ab} }}{\td P_T^2 \td \hat y}
&=&
\frac{\as^2\, y_t^2}{36\,\pi\,\hat{s}^{\,3}\hat{s}_1\,
(\s-\sa+M_H^2)\,\lambda^{5/2}(\hat{s},\hat{s}_1,M_H^2)}
\nonumber\\
&&\times
\bigg\{
C^{ab}(\hat{s},\hat{s}_1,Y)\, \hat{s}\,\hat{s}_1
\,\log\frac{\big[ (\hat{s}-\hat{s}_1+M_H^2) \hat{s}_1^{1/2} 
                        + (\hat{s}_1-4M_t^2)^{1/2}\, \lambda^{1/2}(\hat{s},\hat{s}_1,M_H^2) \big]^2}
               {4\big[M_H^2 \hat{s} \hat{s_1}+M_t^2\lambda(\hat{s},\hat{s}_1,M_H^2)\big]}
\nonumber\\
&&\hspace{1cm}
+D^{ab}(\hat{s},\hat{s}_1,Y)
\hat{s}_1^{1/2} (\hat{s}_1-4M_t^2)^{1/2}\,(\hat{s}-\hat{s}_1+M_H^2)\, \lambda^{1/2}(\s,\sa,M_H^2) \bigg\},
\end{eqnarray}
The functions $C^{ab}$ and $D^{ab}$
are polynomials in the Lorentz invariants $\hat{s}$, $\hat{s}_1$, and $Y$ 
and in the masses $M_t^2$ and $M_H^2$.
They can be expanded in powers of $M_H^2$ as in Eqs.~\eqref{eq:expandMH}.
The expansion coefficients for $C^{\text{ab}}(\hat{s},\hat{s}_1,Y)$ are
\begin{subequations}
\begin{eqnarray}
C^{\text{ab}}_{0} &=&
8M_t^2(\s-\sa)^4(\s+2M_t^2)-64Y M_t^2(\s-\sa)^2(\sa+2M_t^2),
\\
C^{\text{ab}}_{1} &=&
-2(\s-\sa)^2\left[ \s(\s-\sa)^2+2M_t^2(9\s^2-2\s\sa+\sa^2)+16M_t^4(\s+2\sa)\right]
\nonumber\\
&& +16Y\left[ (\s-\sa)^2(\s+2\sa)-2M_t^2(\s^2+6\s\sa-3\sa^2)+16M_t^4 (\s+\sa)\right],
\\
C^{\text{ab}}_{2}&=&
4\left[ \s(\s-\sa)^3+2M_t^2(\s+\sa)(9\s^2-5\s\sa+2\sa^2)+8M_t^4(\s^2+3\sa^2)\right]
\nonumber\\
&& -32Y\left[\s^2+2\sa^2-2M_t^2\s +4M_t^4\right],
\\
C^{\text{ab}}_{3} &=&
-4\left[\s^3-3\s\sa^2+2M_t^2(9\s^2+4\s\sa+3\sa^2)+8M_t^4(\s+2\sa) \right]
\nonumber\\
&& +16Y\left[\s+2\sa-2M_t^2\right],
\\\
C^{\text{ab}}_{4}&=& 4(\s^2-\s\sa+4M_t^2(2\s+\sa)+4M_t^4),
\\
C^{\text{ab}}_{5} &=& -2(s+2M_t^2),
\end{eqnarray}
\end{subequations}
The expansion coefficients for $D^{\text{ab}}(\hat{s},\hat{s}_1,Y)$ are
\begin{subequations}
\begin{eqnarray}
D^{\text{ab}}_{0} &=& \s(\s-\sa)^2\left[ (\s-\sa)^2-8M_t^2\sa\right]
\nonumber\\
&& -8Y\left[ (\s-\sa)^2(\s+\sa)-4M_t^2(\s^2+4\s\sa+\sa^2) \right],
\\
D^{\text{ab}}_{1} &=& -2\s\left[ \s(\s-\sa)^2+4M_t^2(\s^2+2\s\sa-\sa^2)\right]
\nonumber\\
&& +16Y\left[ \s^2-\s\sa+\sa^2-4M_t^2(\s+\sa)\right],
\\
D^{\text{ab}}_{2} &=& 
2\s\left[ \s^2+2\s\sa-\sa^2+4M_t^2(2\s+\sa) \right]-8Y(\s+\sa-4M_t^2),
\\
D^{\text{ab}}_{3} &=&-2\s(\s+4M_t^2),
\\
D^{\text{ab}}_{4}&=&\s.
\end{eqnarray}
\end{subequations}


\providecommand{\href}[2]{#2}\begingroup\raggedright
\endgroup


\end{document}